\documentclass[
reprint
,aps
,amssymb
,amsmath
,superscriptaddress
]{revtex4-1}
\usepackage{graphicx}
\usepackage{dcolumn}
\usepackage{fancyref}
\usepackage{bm}
\usepackage{color}
\usepackage[toc,page]{appendix}
\usepackage[colorlinks=true,allcolors=blue]{hyperref}
\usepackage[normalem]{ulem}

\newcommand{\mdc}[1]{{\color{black}#1}}

\newcommand{\lisa}[1]{{\color{black}#1}}
\newcommand{\dms}[1]{{\color{black}#1}}

\begin{document}

\title{Glassy Dynamics in Models of Confluent Tissue with Mitosis and Apoptosis}

\begin{abstract}
Recent work on particle-based models of tissues has suggested that any finite rate of cell division and cell death is sufficient to fluidize an epithelial tissue. At the same time, experimental evidence has indicated the existence of glassy dynamics in some epithelial layers despite continued cell cycling. To address this discrepancy, we quantify the role of cell birth and death on glassy states in confluent tissues using simulations of an active vertex model that includes cell motility, cell division, and cell death. Our simulation data is consistent with a simple ansatz in which the rate of cell-life cycling and the rate of relaxation of the tissue in the absence of cell cycling contribute independently and additively to the overall rate of cell motion. \mdc{Specifically, we find that a glass-like regime with caging behavior indicated by subdiffusive cell displacements can be achieved in systems with} \dms{sufficiently low rates of cell cycling.}
%Specifically, there is one regime where cell cycling dominates the dynamics and the tissue always fluidizes, and a second glassy regime where the tissue exhibits caging behavior with subdiffusive cell displacements.

\end{abstract}

\author{Michael Czajkowski}
\affiliation{Physics Department, Georgia Institute of Technology}
\author{Daniel M. Sussman}
\affiliation{Physics Department and Soft \& Living Matter Program, Syracuse University}
\author{M. Cristina Marchetti}
\affiliation{Department of Physics, University of California at Santa Barbara}
\author{M. Lisa Manning}
\affiliation{Physics Department and Soft \& Living Matter Program, Syracuse University}
\date{\today}

\maketitle

\section{Introduction}
A number of experiments conducted over the past decade~\cite{Angelini2011,Schotz2013,Park2015} have indicated the existence of glass-like states in confluent epithelial tissues. Evidence for glassy behavior includes the observation of dynamical heterogeneities  characteristic of supercooled liquids~\cite{Angelini2011} and caging effects in motile cell trajectories~\cite{Park2015}. In these living tissues, the constituent cells will often undergo mitosis (cell division) and apoptosis (programmed cell death) in a regulated cell life cycle. These cell life events are necessary for the survival and function of many tissues~\cite{Renehan2001,Thompson1995}, and they clearly differentiate these dense cellular systems from the molecular and polymeric glasses that are so well studied. Introducing and removing material elements from any system should impact the dynamics, and it is therefore important to examine the effect of cell division and cell death on  glassy tissue dynamics.

Recent simulations have suggested that the presence of cell division and death should generically fluidize a tissue~\cite{Matoz-Fernandez2017-1,Matoz-Fernandez2017-2, Ranft2010,Malmi-Kakkada2018}. Employing a particle-based model, for instance, Matoz-Fernandez \textit{et. al.}~\cite{Matoz-Fernandez2017-1} have suggested that, even at small rates $k_\delta$ of cell division and death per cell, one will not observe the sub-diffusive behavior of cell trajectories, which we take to be a hallmark of glassy systems. This, therefore, poses a serious question for the experimental and theoretical studies of these systems: why are caging behavior and other signatures of glassy dynamics observed at all? 

Earlier work by Ranft et. al.~\cite{Ranft2010} at the continuum level considered the effect of cell division and death events on a model of 3-dimensional elastic tissue. Their analytical and numerical findings suggest that cell dynamics will be diffusive and controlled by $k_\delta$ in the limit of long times and infinitely large tissues. However, both this work and many other previous investigations have focused on cell division and death as the dominant driving force behind tissue dynamics. Epithelial cells can also escape the cages set by their neighbors through motility driven by traction forces the cells can exert on a substrate. Such forces play important roles in the development and function of organisms~\cite{Puliafito2012, Martin2004, Trepat2009, Giavazzi2017-1, Giavazzi2017-2}; even in the absence of cell division and cell death, the timescale over which cells may escape their cages changes by orders of magnitude with variation of propulsive forces and intercellular tensions~\cite{Henkes2011, Berthier2013, Berthier2014, Bi2016, Giavazzi2017-2}. The full dynamics of tissues which experience motile forces in addition to mitosis and apoptosis should therefore have contributions from each of these sources of internal driving. In particular, it is natural to expect that the observed relaxation time of a tissue will be set by a competition between the timescale of cell cycling events and the timescale of cage escape due to the motile cells encountering mechanical energy barriers.

A complicating factor in previous investigations of model tissues with both active dynamics and cell cycle events is that they have focused on particle-based modes and they often couple the details of the cell cycle events to the current configurational state of the tissue~\cite{Matoz-Fernandez2017-1,Matoz-Fernandez2017-2, Ranft2010} (for instance, triggering cell divisions when a particle is surrounded by a sufficient amount of space, or triggering cell deaths when the local density or contact number is too high). While such couplings may be biologically motivated, they also make it difficult to separate out the mechanically fluidizing effects of cell division and death on the glassy behavior of the model tissues. 

In this work, we focus on models of confluent tissues, where there are no gaps or overlaps between cells.  Such tissues are well-described by so-called ``vertex models'' that have been used extensively to understand patterning and rigidity in tissues~\cite{Farhadifar2007,Hufnagel2007,Staple2010, Bi2014, Bi2015, Merkel2017}. Although there has been substantial recent work on ``Voronoi'' versions of such models~\cite{Bi2016,Barton2017}, here we introduce an ``Active Vertex Model'' (AVM) which naturally incorporates cell motile forces similar to the previously studied Self-Propelled Voronoi (SPV) model, but without the restriction that cell shapes be described by Voronoi volumes.

The remainder of this work is organized as follows. In Sec. \ref{sec:AVM} we describe our Active Vertex Model and characterize its dynamical behavior in the absence of cell division and cell death. In Sec. \ref{sec:dd} we extend this model to include a single controlled rate of mitosis and apoptosis, and we demonstrate that the phase diagram for this model has two dynamical regimes: \mdc{one where the system is dominated by cell cycle events and another where the system is controlled by the inherent dynamics of the motility-only model.} %\mdc{TWO REGIME NOTE} two dynamical regimes: a fluid-like regime dominated by cell cycle events and a glassy regime dominated by the inherent slow dynamics of the \mdc{motility-only} model. 
These combined regimes are well-described by a simple ansatz in which the overall relaxation time is an additive independent sum of the inherent rate of cell motility and the rate of cell cycling, revealing a universal crossover between cell-cycle-dominated and motility-dominated dynamics. In Sec. \ref{sec:dd_single_events} we investigate the displacement fields associated with individual mitosis and apoptosis events. Each event produces an Eshelby-like displacement field of tissue cells, in addition to a mechanical noise stemming from the disordered geometry of the tissue. Quantifying these displacements allows us to construct a prediction for the cell diffusion across a broad range of cell cycling rates, confirming that the effect of cell cycling on diffusion is similar to that predicted by previous continuum models~\cite{Matoz-Fernandez2017-1,Ranft2010}, although the \mdc{fluctuating} motions generated by disordered geometries that are typically ignored by those models also contribute to overall cell motion. We close in Sec. \ref{sec:discussion} with a discussion of how our results compare to previous work, and we suggest experiments to guide future developments.

\section{A Vertex Model for Motile Tissues\label{sec:AVM}}

Numerical studies of the influence of motility on tissue dynamics have been conducted in particulate models~\cite{Henkes2011}, the Potts model~\cite{Kabla2012, Chiang2016}, and Voronoi / vertex models~\cite{Bi2016, Giavazzi2017-2,Sussman2017-2,Sussman2018-2}. These investigations have generally indicated that motile forces tend to promote fluidity and enhance diffusion. For instance, using the SPV model -- in which cells are described as polygons obtained from a Voronoi tiling of the plane that self-propel analogously with self-propelled particle models -- Bi \textit{et. al.}~\cite{Bi2016} obtained a transition line separating glassy states from fluid-like states where cells frequently exchange neighbors. %Notably, this same investigation revealed that this transition is described by an order parameter based on cell shapes. Some experimental evidence has already started to confirm  a dimensionless ``shape index'' may be sufficient to predict whether cells will migrate through real tissues~\cite{Park2015}.
%The numerical investigation which revealed this transition was performed using the SPV model.
%In the SPV model, cells are described as polygons obtained from a Voronoi tiling of the plane.

While the Voronoi description is theoretically appealing, the restriction of cell shapes to local Voronoi volumes is not always desirable. More seriously, Voronoi models are not easily extendable to describe tissues with free boundaries, as would be needed to describe wound-healing geometries in 2D or isolated spheroids in 3D. To avoid these issues, while simultaneously preserving the spirit of the self-propelled Voronoi model, we have constructed a model for a motile tissue that preserves the freedom of polygonal cell shapes. In this work we restrict our attention to the generic case where every vertex is three-fold coordinated, although this restriction could be easily relaxed to study the sorts of open boundaries described above. To our knowledge, the dynamics of the Active Vertex Model presented herein has not been explored in previous work. Therefore, we include a more detailed description of the model, as well as quantification of its dynamical states in the absence of cell cycle events.

%% While the Voronoi description is simple and fast to simulate, a recent study by Sussman and Merkel has indicated that in two dimensions the Voronoi tiling constraint may have a substantial impact on tissue rigidity~\cite{Sussman2017-2}. Specifically, the 2D Voronoi model has very strange low temperature behavior due to the fact that the number of linear constraints exactly equal the number of degrees of freedom.  To avoid these issues we have constructed a model for a motile tissue that preserves the freedom of polygonal cell shapes. To our knowledge, the ``Active Vertex Model'' presented herein has not been explored in any previous work. Therefore, we include a more detailed description of the model, as well as quantification of the solid-like and fluid-like states.

\begin{figure*}[!t]  
\begin{center}
    \includegraphics[width=0.7\textwidth]{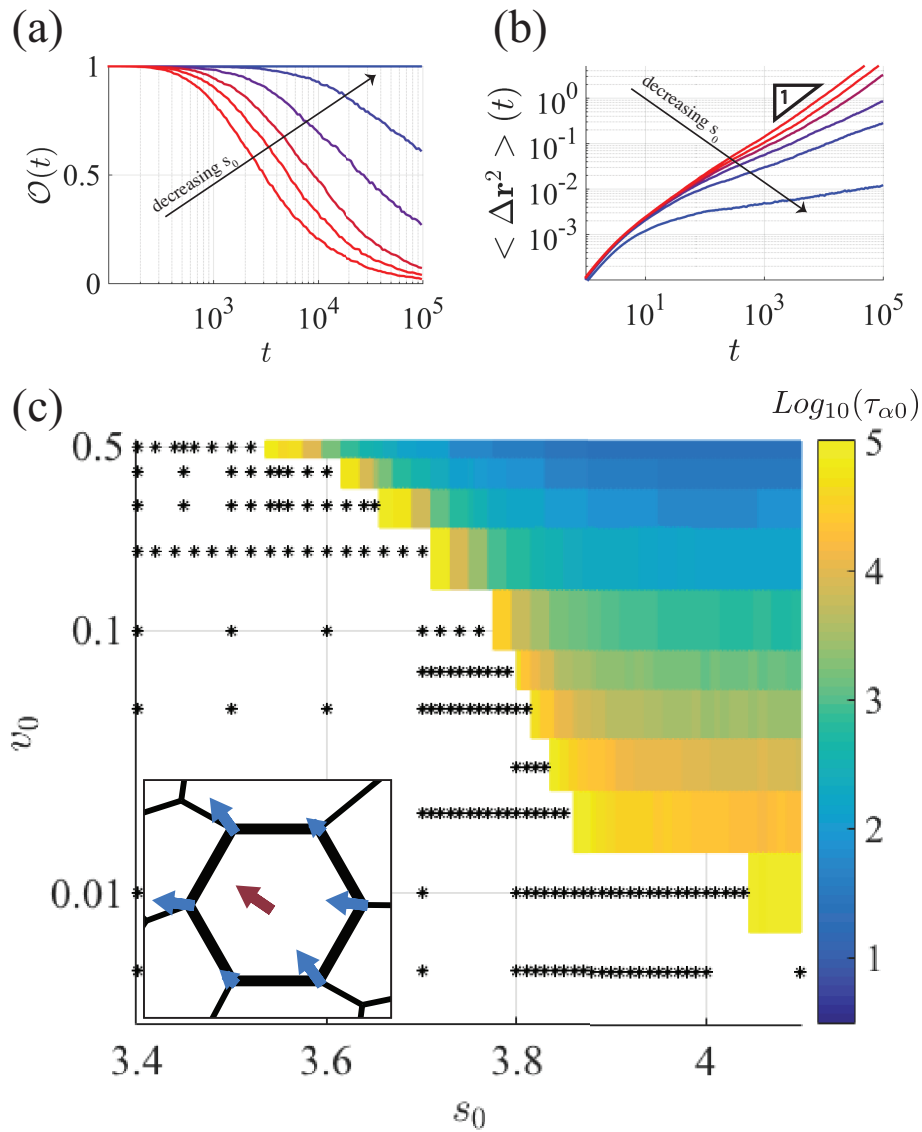}
    \caption{\label{fig:AVM_phase_behavior} Quantification of phase behavior in the Active Vertex Model (AVM) with $D_r=1$ and $N_{cells}=300$. In (a), the timescale $\tau_{\alpha0}$ of the decay of the self-overlap function grows beyond the length of simulations as the vertex model tuning parameter $s_0$ is decreased while $v_0=0.05$ is held fixed. In (b), the mean square displacement of cell centers for the same phase points indicates this increasing timescale is associated with subdiffusive caging behavior. In (c), $\tau_{\alpha0}$ is shown on a logarithmic colorscale for a representative set of points in the $s_0-v_0$ plane. Black stars indicate phase points where $\tau_{\alpha0}$ is too long to be resolved with our data. The inset shows the implementation of \mdc{ cell motility in terms of vertex forces as prescribed by } Eq.~\eqref{eq:AVM_total_energy}. }
    \end{center}
\end{figure*}

%%%%%%%%%%%%%%%%%%%%%%%%%%%%%%%%%%%%%%%%%%%%%%%%%%%%%%%%%%%%%%%%%%%%%%%%%%%%%%%%%%%%%%
%%%%%%%%%%%%%%%%%%%%%%%%%%%%%%%%%%%%%%%%%%%%%%%%%%%%%%%%%%%%%%%%%%%%%%%%%%%%%%%%%%%%%%
%%%%%%%%%%%%%%%%%%%%%%%%%%%%%%%%%%%%%%%%%%%%%%%%%%%%%%%%%%%%%%%%%%%%%%%%%%%%%%%%%%%%%%

\subsection{Shape forces in the Vertex Model}
In the Vertex Model~\cite{Farhadifar2007, Staple2010, Hufnagel2007, Bi2014} cells are modeled as irregular polygons tiling the plane, but in contrast to Voronoi models, the degrees of freedom are the positions of the vertices of the spatial tiling. Previous studies using this model have typically involved searching for geometric states which minimize the tissue energy,
\begin{equation}\label{eq:AVM_evertex}
E_{Shape} = \sum_{a} E_a = \sum_{a} \left[ \kappa_{A}\left(A_a-A_{0a}\right)^2+\kappa_{P}\left(P_{a}-P_{0a}\right)^2 \right] \, ,
\end{equation}
where the sum runs over all cells $a$ and $\kappa_A$ and $\kappa_P$ are elastic moduli. This energy drives the area $A_a$ and perimeter $P_a$ of cell $a$ toward a target area $A_{0a}$ and a target perimeter $P_{0a}$ respectively. The area term in Eq.~\ref{eq:AVM_evertex} comes from the resistance to fluctuations in cell height. The perimeter term may be seen as arising from the competition between cell-cell adhesive interactions and the non-linear tension created by the cortical actin network. While the target shape properties may vary from cell to cell, in the absence of cell division and death we presume cells identical with uniform $A_0$ and $P_0$ values. An important parameter in such models is the dimensionless target ``shape index'' $s_0 = P_0 / \sqrt{A_0}$, which controls rigidity in both static~\cite{Bi2015} and motile~\cite{Bi2016} tissues.

%%%%%%%%%%%%%%%%%%%%%%%%%%%%%%%%%%%%%%%%%%%%%%%%%%%%%%%%%%%%%%%%%%%%%%%%%%%%%%%%%%%%%%
%%%%%%%%%%%%%%%%%%%%%%%%%%%%%%%%%%%%%%%%%%%%%%%%%%%%%%%%%%%%%%%%%%%%%%%%%%%%%%%%%%%%%%
%%%%%%%%%%%%%%%%%%%%%%%%%%%%%%%%%%%%%%%%%%%%%%%%%%%%%%%%%%%%%%%%%%%%%%%%%%%%%%%%%%%%%%

\subsection{The Active Vertex Model}
Although the simplest implementation of activity in a vertex model allows each vertex to be self-propelled, we would like to model the behavior of a polarized motile cell moving persistently along a specific direction. Therefore, we extend this model to include self propulsion of cell $a$ in the direction
\begin{equation}\label{eq:AVM_nhat}
\mathbf{\hat{n}}(\theta_a) = \cos(\theta_a)\mathbf{\hat{x}} + \sin(\theta_a)\mathbf{\hat{y}} \, ,
\end{equation}
where the cell propulsion angle $\theta_a$ is governed by
\begin{equation}\label{eq:AVM_theta_dynamics}
\partial_t \theta_a = \eta_a \, ,
\end{equation}
with $\eta_a$ being a white gaussian noise defined by 
\begin{equation}\label{eq:AVM_noise_average}
<\eta_a(t)> = 0\, 
\end{equation}
and
\begin{equation}\label{eq:AVM_noise_variance}
<\eta_a(t_1)\eta_{a'}(t_2)> = 2 D_r \delta(t_1 - t_2) \delta_{aa'} \, .
\end{equation}
Assuming that cell dynamics take place in the overdamped limit, vertex $\mu$ will follow an equation of motion,
\begin{equation}\label{eq:AVM_vert_dynamics}
\partial_t \mathbf{r}^\mu = -\frac{1}{\gamma} \frac{\partial}{\partial\mathbf{r}^\mu} E_{AVM} \, ,
\end{equation}
where $\gamma$ is the substrate friction. Here and throughout we use superscripted greek letters to refer to vertex positions, and subscripted latin letters to refer to cell positions. The total effective energy,
\begin{equation}\label{eq:AVM_total_energy}
E_{AVM} = E_{Shape}  - \gamma v_0 \sum_{cells-a}  \mathbf{\hat{n}}(\theta_a) \cdot \mathbf{r}_a \, ,
\end{equation}
now captures the cellular self-propulsion forces (of magnitude $v_0$) in addition to the standard shape energy from Eq.~\ref{eq:AVM_evertex}. In our AVM simulations we define $\mathbf{r}^a$ as the cell centroid, or geometric center, 
\begin{equation}\label{eq:centroid}
\mathbf{r}_a = \frac{1}{6 A_a}  \sum_{\nu} \left( \mathbf{r}_a^\nu  + \mathbf{r}_a^{\nu+1} \right) (\mathbf{r}_a^\nu  \times \mathbf{r}_a^{\nu+1}) \cdot \mathbf{\hat{k}} \, ,
\end{equation}
where $\nu$ here indexes the $N_a$ vertices on cell $a$ in counterclockwise fashion and $\mathbf{r}_a^{N_a + 1} = \mathbf{r}_a^1$.
%the sum goes counterclockwise over the $N_a$ vertices on cell $a$ and $\mathbf{r}_a^{N_a + 1} = \mathbf{r}_a^1$.

%In this model, 
\mdc{With these equations of motion, each cell attempts to carry out a persistent random walk, as is commonly observed for cells isolated from the surrounding tissue and in the absence of external signals~\cite{Metzner2015, Passucci2019}.} While it is uncommon to think of such a self-propulsion force as coming from an energy~\cite{sknepnek2015active}, writing things in this way permits an analogy with the Voronoi model. In the SPV, the degrees of freedom are the Voronoi centers and we might imagine constructing the same energy Eq.~\ref{eq:AVM_total_energy}, using these Voronoi centers as the $\{\mathbf{r}_a\}$. It is easy to see that using these as the degrees of freedom in the overdamped equation of motion,
\begin{equation}
\begin{aligned}\label{eq:AVM_voronoi_dof}
\partial_t \mathbf{r}_a & = -\frac{1}{\gamma} \frac{\partial}{\partial\mathbf{r}_a} E \\ &= v_0 \mathbf{\hat{n}}(\theta_a) -\frac{1}{\gamma} \frac{\partial}{\partial\mathbf{r}_a} E_{shape} \, ,
\end{aligned}
\end{equation}
indeed leads to the standard SPV dynamics~\cite{Bi2016,Yang2017,Barton2017}. We are therefore using the closest energetic analog that implements SPV dynamics in a vertex model. In the AVM we will instead apply shape-based and motility-based forces to the vertices. These forces may be precisely defined by taking the derivatives in Eq.~\ref{eq:AVM_vert_dynamics}, as explained in greater depth in Appendix~\ref{app:avm_explained}.

%%%%%%%%%%%%%%%%%%%%%%%%%%%%%%%%%%%%%%%%%%%%%%%%%%%%%%%%%%%%%%%%%%%%%%%%%%%%%%%%%%%%%%
%%%%%%%%%%%%%%%%%%%%%%%%%%%%%%%%%%%%%%%%%%%%%%%%%%%%%%%%%%%%%%%%%%%%%%%%%%%%%%%%%%%%%%
%%%%%%%%%%%%%%%%%%%%%%%%%%%%%%%%%%%%%%%%%%%%%%%%%%%%%%%%%%%%%%%%%%%%%%%%%%%%%%%%%%%%%%

\subsection{Glassy dynamics in the Active Vertex Model \label{sec:AVM_rheology}}
Since this model for confluent tissue dynamics has not been explored in previous work, we first examine its behavior and search for glassy states. We quantify the dynamical state of the system by using two familiar metrics: the mean square displacement and the self-overlap function. The self-overlap function is defined as
\begin{equation}\label{eq:overlap_definition}
\mathcal{O}(t) = \frac{1}{N_{cells}} \sum_{a} \Theta\left( b-| \mathbf{r}_a(t) -\mathbf{r}_a(0)|\right)  \, ,
\end{equation}
%
% phase diagram using the techniques described in Bi et. al.~\cite{Bi2016}. 
where $\Theta$ is the heaviside function and $b$ represents the size of a typical cage in natural units, which we set to $0.5$. All lengths are measured in units of $1/\sqrt{\rho}$, where $\rho$ is the system number density. The function $\mathcal{O}$ has a value of $1$ at $t=0$ and decays towards zero as cells move beyond the caging distance $b$. The structural relaxation time $\tau_{\alpha0}$ is defined as the point at which the self-overlap decays below $1/e$.

As indicated in the introduction, in this work we use the term ``glassy'' to primarily refer to the presence of a subdiffusive regime in the mean squared displacements, and to the observation of uncaging times $\tau_{\alpha0}$ that are orders of magnitude longer than expected from free diffusion. %We search for these signatures of supercooled fluids to establish expectations for the AVM, with a particular focus on the uncaging time. 
The uncaging time averaged over 10 systems of $N_{cells}=300$ is displayed in Fig.~\ref{fig:AVM_phase_behavior}. Its behavior is consistent with previous work on the Self-Propelled~\cite{Bi2016} and Thermal~\cite{Sussman2018-2} Voronoi models. Less motile tissues at lower $s_0$ (higher cortical tension) are indeed glassy and have long relaxation times. The tissue may be effectively fluidized by either increasing $v_0$ (higher effective temperature) or increasing $s_0$ (higher cell-cell adhesion). 

Although the overlap function shown in Fig.~\ref{fig:AVM_phase_behavior} appears qualitatively similar to those observed in other glassy materials, %a very interesting and important point is that many of 
both it and the mean-squared displacement shows an unusual behavior.  In most glassy materials, the mean-squared displacement levels off to a subdiffusive regime where $\Delta r^2 \sim t^0$,  followed later by a diffusive regime $\Delta r^2 \sim t^1$.  Here we find anomalous sub-diffusive behavior with a non-standard exponent, $\Delta r^2 \sim t^\alpha$, where $0< \alpha < 1$, is found to persist across a wide range of timescales. In this same parameter regime, we also see multiple timescales in the shear stress autocorrelation function. These results are highlighted in Appendix~\ref{app:avm_rheology}.  Such properties are highly atypical in the context of standard glasses, and may be related to the unusual zero-temperature rigidity transition in such models \cite{Sussman2017-2}. Understanding the origins of this behavior will be an interesting avenue for future work.

As we are interested in characterizing the simple long-time diffusion of cells in our model, this unusual behavior is problematic. To acknowledge this, throughout the manuscript we highlight with distinct symbols any data presented from simulations that remain sub-diffusive on long timescales. In Section~\ref{sec:dd} we will see that the parameter regimes of the active vertex model that are diffusive at long times have dynamics which are easily interpreted as a competition between glassy relaxation and relaxation driven by cell division and death.

\section{Simulation of mitosis and apoptosis events \label{sec:dd}}

We would now like to incorporate the influence of cell cycle events in these vertex model simulations. To identify an appropriate procedure, we first note that division and death events can change the total number of cells and the cell areal density, with potentially drastic changes in tissue dynamics. 
% acknowledge the expected influence of changing cell number in our model. Consider a tissue of cells having some target perimeter $P_0$ and target area $A_0$ and for which each cell is able to realize these target values. If some apoptosis events occur with the system size fixed, then these cells will no longer be able to realize their target $A_0$ as they must continue to tile the same space. Lending from known results from the rigidity transition in these tissues~\cite{Bi2015, Merkel2018} we might also expect that the cells will no longer be able to realize their target $P_0$, and thereby a potentially drastic change in tissue rheology may result. 
To minimize such effects, we choose to work in a constant-number ensemble. In our simple implementation, each instance of apoptosis will be accompanied by a mitosis event somewhere else in the tissue. In addition, the implementation of the individual cell cycle events is chosen to preserve the sum of the cell target areas.% $A_{0}^{(T)}= \sum_{a}A_{0a}$ at all times.

\begin{figure}[!t]  
\begin{center}
    \includegraphics[width=0.48\textwidth]{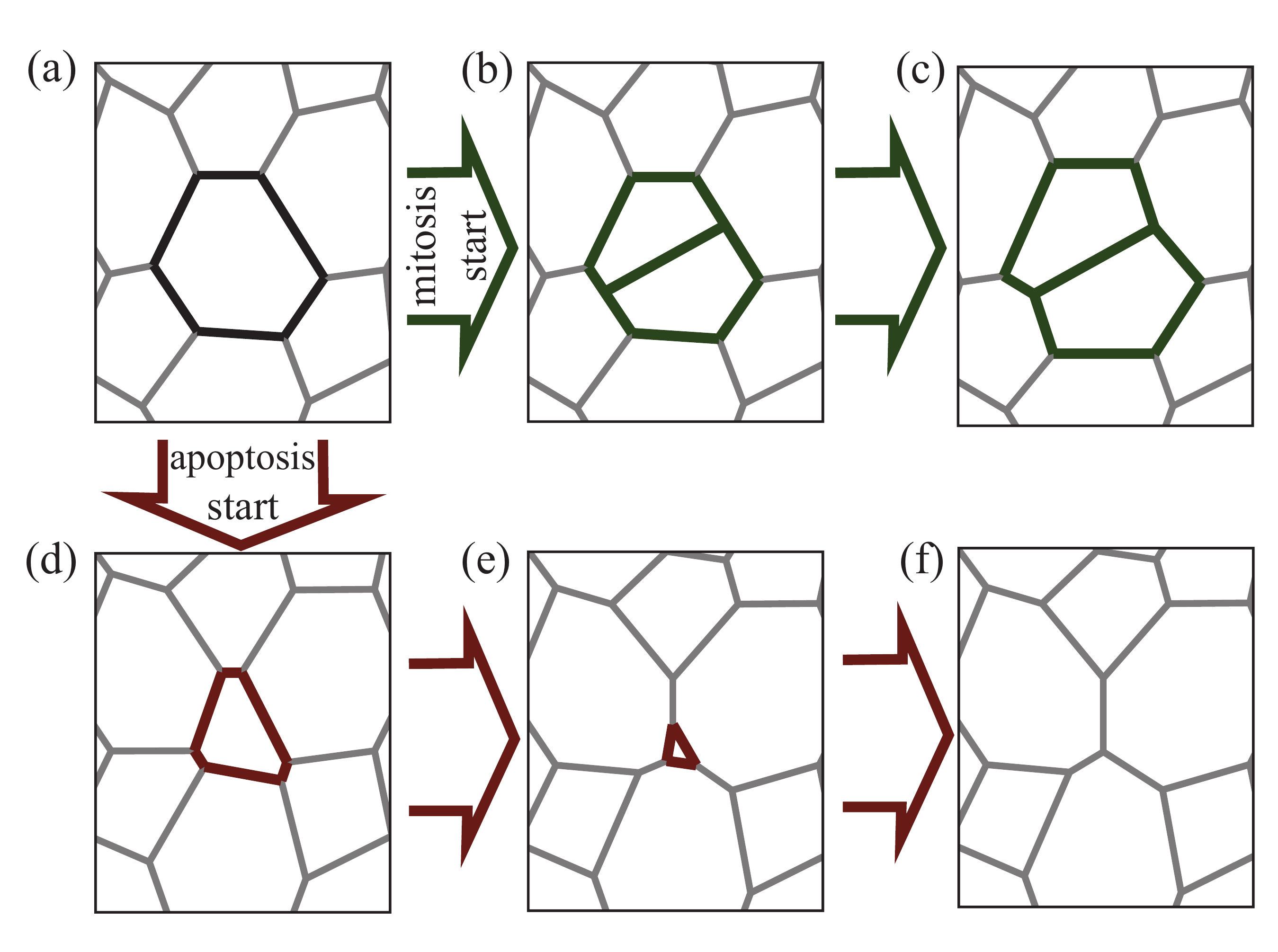}
    \caption{\label{fig:dd_birth_death_cartoon} The process for a single mitosis event in our model is shown sequentially in (a-c). Starting again with the initial cell in (a), the model steps for apoptosis are shown sequentially in (d-f).}
    \end{center}
\end{figure}

In apoptosis, a cell will abruptly contract to a small size and then extrude itself, effectively disappearing from the 2d monolayer~\cite{Rosenblatt2001}. Shown in Fig.\ref{fig:dd_birth_death_cartoon}(d-f), our simple realization of apoptosis on cell $a$ begins with setting the target area and perimeter to zero. %$A_{0a} \rightarrow 0$ and $P_{0a} \rightarrow 0$. 
This change induces the rapid contraction of cell $a$ to small size and generally will lead to a final triangular shape. The simulation then detects triangular cells that are smaller than some threshold area $A_{min}$ and deletes them.

Similarly, in the process of mitosis a cell will expand, eventually reaching a threshold size and dividing into two complete cells~\cite{Conlon2003,Dolznig2004,Streichan2014}, all the while maintaining tissue cohesion~\cite{Baker1993}. While this process typically spans a much longer time period than the event of a cell death, we model this growth process as similarly instantaneous to suppress fluctuations of tissue density. %preserve the constant $A_{0T}$. 
Shown in Fig~\ref{fig:dd_birth_death_cartoon}, in the division of cell $a$ two non-adjacent edges of the cell are chosen at random, and their midpoints determine the axis of the division. A vertex is then added to the center of each edge and these new vertices are connected with a new edge. Naturally the product is two closed polygonal cells out of one, and the shape parameters of these new cells are then set to the values of the parent cell. These cells are then allowed to expand dynamically in the simulation. Similar division dynamics were studied in a quasi-static system by others, including Farhadifar \textit{et. al.}~\cite{Farhadifar2007}.

Such constant number simulations allow us to enforce a global cell cycle event rate $k_\Delta$, which is implemented as a Poisson process. This sets the effective cycle rate per cell as $k_\delta = k_\Delta / N_{cells}$. Note that $k_\delta$ is a more appropriate parameter for real biological systems whose cell cycle timing does not depend strongly on the extent of the surrounding tissue. We therefore use $k_\delta = 1/\tau_\delta$ as our tuning parameter.

In simulations with cell division and death, complications arise when trying to extract cell trajectory data. This is because many trajectories will start and end during the course of the simulation. Rather than parse this network of trajectories, we follow the method used in~\cite{Matoz-Fernandez2017-1} and exclude a small (\%10) subset of the cells from the cell cycle events. Using these ``tracer'' cells to obtain our dynamical data is both convenient and affords the closest comparison with the previous literature~\cite{Matoz-Fernandez2017-1}. 

We note that fast rates of cell division tend to cause tissues to reach rare topological states that \mdc{can generate a breakdown in our simulation framework;} %introduce very stiff coefficients in the ODEs and are not easily handled by our numerical integrator\dms{MICHAEL is checking this}. 
to avoid this we restrict the data presented here to $\tau_\delta > 3000$.% to avoid any bias from missing simulation data. 

%%%%%%%%%%%%%%%%%%%%%%%%%%%%%%%%%%%%%%%%%%%%%%%%%%%%%%%%%%%%%%%%%%%%%%%%%%%%%%%%%%%%%%
%%%%%%%%%%%%%%%%%%%%%%%%%%%%%%%%%%%%%%%%%%%%%%%%%%%%%%%%%%%%%%%%%%%%%%%%%%%%%%%%%%%%%%
%%%%%%%%%%%%%%%%%%%%%%%%%%%%%%%%%%%%%%%%%%%%%%%%%%%%%%%%%%%%%%%%%%%%%%%%%%%%%%%%%%%%%%

\subsection{Tissue dynamics in the presence of mitosis and apoptosis}

Given this simulation protocol, we search for signatures of glassy behavior. Tissue dynamics is quantified using the uncaging time $\tau_\alpha$, defined as the time required for the observed self-overlap to decay below $1/e$. The notation intentionally differs from that in Section~\ref{sec:AVM_rheology}, to distinguish $\tau_{\alpha0}$ as the uncaging time in the limit of the model without cell cycle events (i.e., the limit $k_\delta=0$).

\begin{figure*}[!t]  
\begin{center}
    \includegraphics[width=0.7\textwidth]{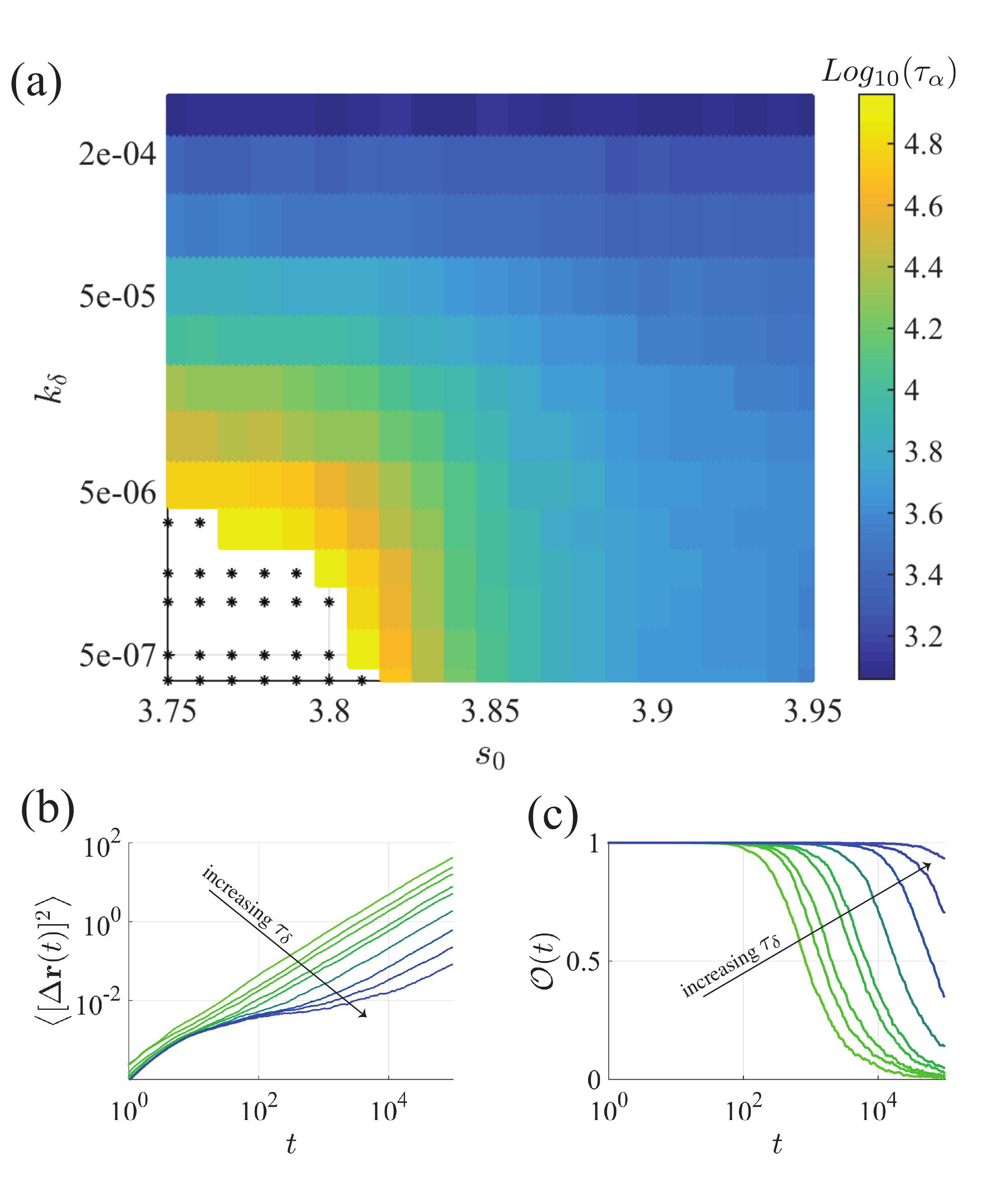}
    \caption{\label{fig:dd_rheology} Dynamical measures of tissues with mitosis and apoptosis for $v_0=0.05$, $D_r=1$, and $N_{cells} = 300$. In (a), the measured overlap decay $\tau_\alpha$ as a function of $s_0$ and $k_\delta$ (the cell cycling rate) is displayed as a colormap, indicating generically that $k_\delta$ tends to fluidize the tissue.  Black stars indicate where $\tau_\alpha$ is beyond the length of our simulations. In (b) and (c), mean square displacement and overlap curves are plotted for constant $s_0=3.76$ over a series (from green to blue) of $\tau_\delta = 3k, 6k, 9k, 20k, 30k, 90k, 300k, 900k, 3M$.% \mdc{The following sentence will need to be removed} These characteristic examples show that the dynamics approaches the expected values from the ``bare'' AVM at long enough $\tau_\delta$. %In (d) and (e) are a similar series of curves for a constant $k_\delta = 1.\bar{6}e-5$ over a series of $s_0$ values. Here, the approach towards a universal behavior (expected at the lowest values of $s_0$) only becomes clear at long times.
    }
    \end{center}
\end{figure*}

Our search for glassy behavior is guided by the simple expectation that the dynamics is determined by a competition between the timescale $\tau_\delta$ of division and death events and the timescale of motility driven cage escape $\tau_{\alpha0}$. Naturally, we recover the dynamics of the ``bare'' (free of cell division and death) AVM in the limit $\tau_\delta \gg \tau_{\alpha0}$ where mitosis and apoptosis will play a negligible role. %for sufficiently large (slow) values of $\tau_\delta$. %\mdc{the following two sentences may need to be removed} This can be seen in Fig.~\ref{fig:dd_rheology} (b) and (c), where values of $k_\delta$ leads the msd and the self-overlap to approach a limiting form which matches the data from Fig.~\ref{fig:AVM_phase_behavior}. This is useful for identifying the limits of when cell division and cell death events have an impact on mechanics.
%That the influence of mitosis and apoptosis will be negligible in the limit $\tau_\delta \gg \tau_{\alpha0}$ where motility operates on a much faster timescale, is reasonable. 
One may then expect to find the opposite behavior in the limit $\tau_\delta \ll \tau_{\alpha0}$, where motility-based dynamics should become negligible. \mdc{To check this, we increase $s_0$ to decrease $\tau_{\alpha0}$. In the top-left region of Fig.~\ref{fig:dd_rheology}-a where} $\tau_{\alpha0}$ is large and $\tau_\delta$ is small ($k_\delta$ is high) we observe structural relaxation times $\tau_\alpha$ which are approximately independent of $s_0$ and by proxy, $\tau_{\alpha0}$.

%%%%%%%%%%%%%%%%%%%%%%%%%%%%%%%%%%%%%%%%%%%%%%%%%%%%%%%%%%%%%%%%%%%%%%%%%%%%%%%%%%%%%%
%%%%%%%%%%%%%%%%%%%%%%%%%%%%%%%%%%%%%%%%%%%%%%%%%%%%%%%%%%%%%%%%%%%%%%%%%%%%%%%%%%%%%%
%%%%%%%%%%%%%%%%%%%%%%%%%%%%%%%%%%%%%%%%%%%%%%%%%%%%%%%%%%%%%%%%%%%%%%%%%%%%%%%%%%%%%%

\begin{figure*}[!t]  
\begin{center}
    \includegraphics[width=0.7\textwidth]{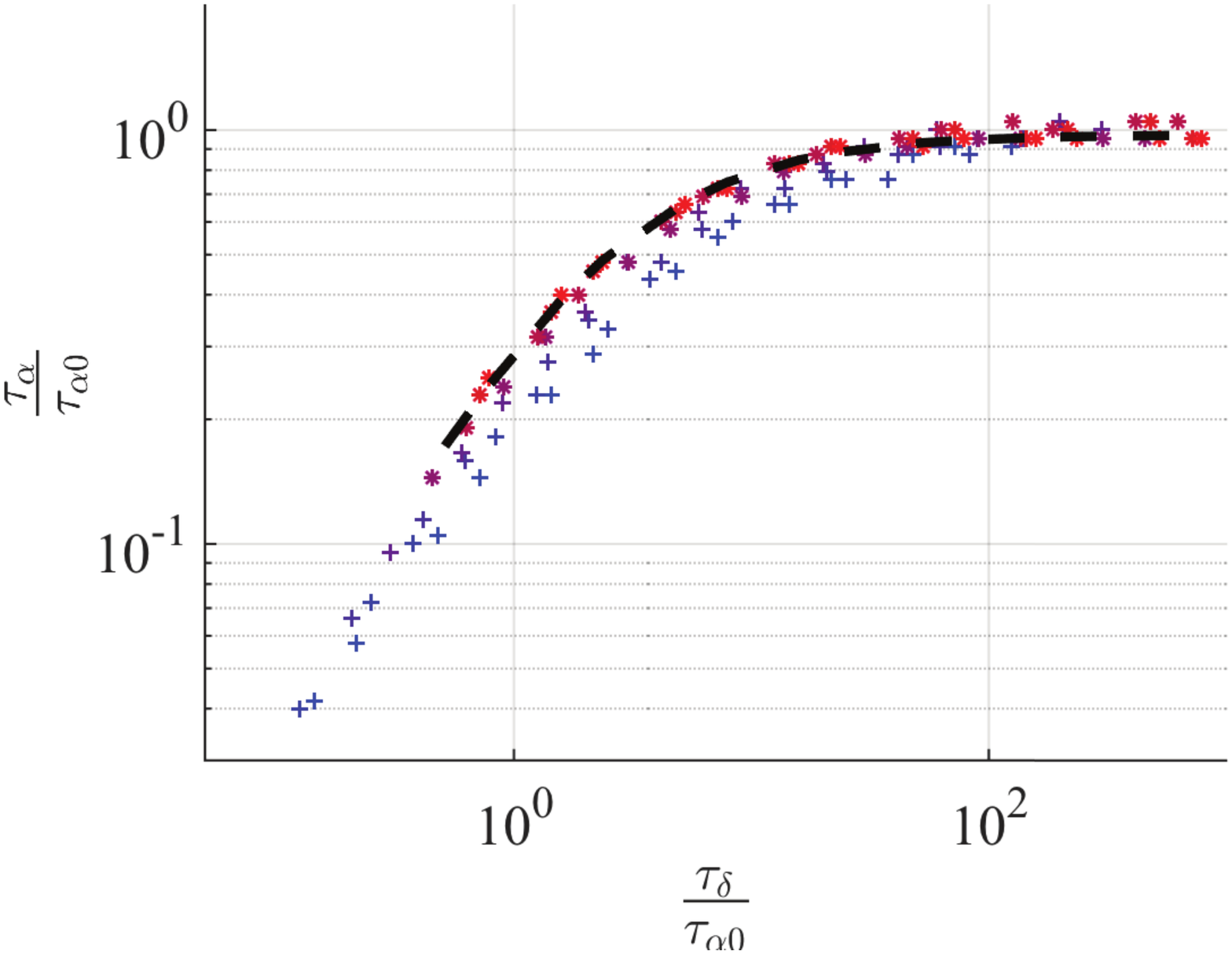}
    \caption{\label{fig:dd_collapses} Observed uncaging times for the AVM with division and death at a range of rates $\tau_\delta$ and (from blue to red) $s_0 = (3.82, 3.825, 3.83, 3.84, 3.85, 3.875, 3.9, 3.925, 3.95)$. The $\tau_\alpha$ values come close to collapse with simple rescaling using $\tau_{\alpha0}$ and suggest a universal crossover between motility-dominated and division-death dominated dynamics. Data is represented by ``+" symbol where diffusion is never observed in the corresponding ``bare'' AVM and is represented by ``*" otherwise. The black dashed line represents the prediction via Eq.~\ref{eq:dd_taualpha1} for $s_0=3.95$, which matches well with the simulation data. All data is from simulations with $N_{cells} = 3000$, $D_r=1.0$ and $v_0 = 0.05$.  
    }
    \end{center}
\end{figure*}

\subsection{Simple model for interaction between \texorpdfstring{$\tau_\delta$}{Td} and \texorpdfstring{$\tau_{\alpha0}$}{Ta0}}

Building on the limiting behaviors described above % agree with expectations, and hint at possibilities for understanding data in the intermediate regime where both timescales $\tau_{\alpha0}$ and $\tau_\delta$ will play a role. To establish expectations in this regime, 
we develop a very simple ansatz for the interplay between cell death and division and glassy dynamics. Specifically, we assume that the overlap decay rate $\frac{1}{\tau_\alpha}$ of a dividing tissue is determined by the weighted sum of the bare cage escape rate $\frac{1}{\tau_{\alpha0}}$ and the rate of division and death events according to
\begin{equation}\label{eq:dd_taualpha1}
\frac{1}{\tau_\alpha} =   \frac{1}{\tau_{\alpha0}} + \frac{C_2}{\tau_\delta}  \, ,
\end{equation}
where $C_2$ captures the displacements of surrounding cells resulting from the division and death events. %The weight $C_1$, which captures the relative contribution of bare glassy dynamics, is set to unity based on the behavior identified in the previous section for the $\tau_\delta \rightarrow \intfy$ limit. 
The strong assumption we have made here is that the two rates add in series and are not strongly correlated.
Our model predicts that the quantity $\frac{\tau_\alpha}{\tau_{\alpha0}}$ will be a function only of  $\frac{\tau_{\delta}}{ \tau_{\alpha0}}$. 
As shown in Fig.~\ref{fig:dd_collapses}, this works reasonably well, indicating the generic presence of a regime where fast divisions dominate the dynamics and $\tau_\alpha$ is proportional to $\tau_\delta$ (left side of Fig.~\ref{fig:dd_collapses}), along with a regime of slow divisions where $\tau_\alpha $ becomes independent of $ \tau_\delta$ (right side of Fig.~\ref{fig:dd_collapses}). To understand what controls the crossover, we quantify the impact of cell division and cell death events on the motion of surrounding cells in the next section. 

%However, there is not a perfect collapse, and the slower, more glassy points tend to display increasing deviation. Among the possible reasons for this, we note both that the factor $C_2$ may depend strongly on glassiness through the parameter $s_0$, and in addition, that there may be correlations between motility-driven cage breaking events and the spatiotemporal distribution of cell cycling events. We begin to address these possibilities in the next sections.
%One possibility is that the coefficient $C_2$, which captures displacements generated by cell division and death, does depend significantly on the inherent tissue rheology $C_2=C_2(\tau_{\alpha0})$. This possibility is explored further in the next section.

%%%%%%%%%%%%%%%%%%%%%%%%%%%%%%%%%%%%%%%%%%%%%%%%%%%%%%%%%%%%%%%%%%%%%%%%%%%%%%%%%%%%%%
%%%%%%%%%%%%%%%%%%%%%%%%%%%%%%%%%%%%%%%%%%%%%%%%%%%%%%%%%%%%%%%%%%%%%%%%%%%%%%%%%%%%%%
%%%%%%%%%%%%%%%%%%%%%%%%%%%%%%%%%%%%%%%%%%%%%%%%%%%%%%%%%%%%%%%%%%%%%%%%%%%%%%%%%%%%%%

\section{Flow and fluidization from individual mitosis and apoptosis events \label{sec:dd_single_events}}

To quantify the effect of a single cell cycling (division or death) event on the motion of the surrounding tissue, we perform a special set of ``single-event'' simulations. In these, the standard AVM is run for a short equilibration time before a single cell is chosen at random to undergo either apoptosis or mitosis in a randomly chosen direction. We then monitor the motion of the surrounding tissue cells and average over 1000 realizations. 

We measure the individual cell displacement vectors $\{ \mathbf{u}_i \}$ and use them to construct a few useful quantities. In order to resolve coherent spatial data, the cells are first binned based on their distance from the event (and in the case of cell division based on their angle relative to the division axis orientation). Within each bin, we calculate two quantities: the vector averaged displacement $\mathbf{u}_{(m,a)}(r, \theta)$ and the vector standard deviation of this average, $\bm{w}_{(m,a)}(r, \theta)$, where the subscripts $m$ and $a$ refer to mitosis and apoptosis events, respectively. 

\begin{figure*}[!t]  
\begin{center}
    \includegraphics[width=0.8\textwidth]{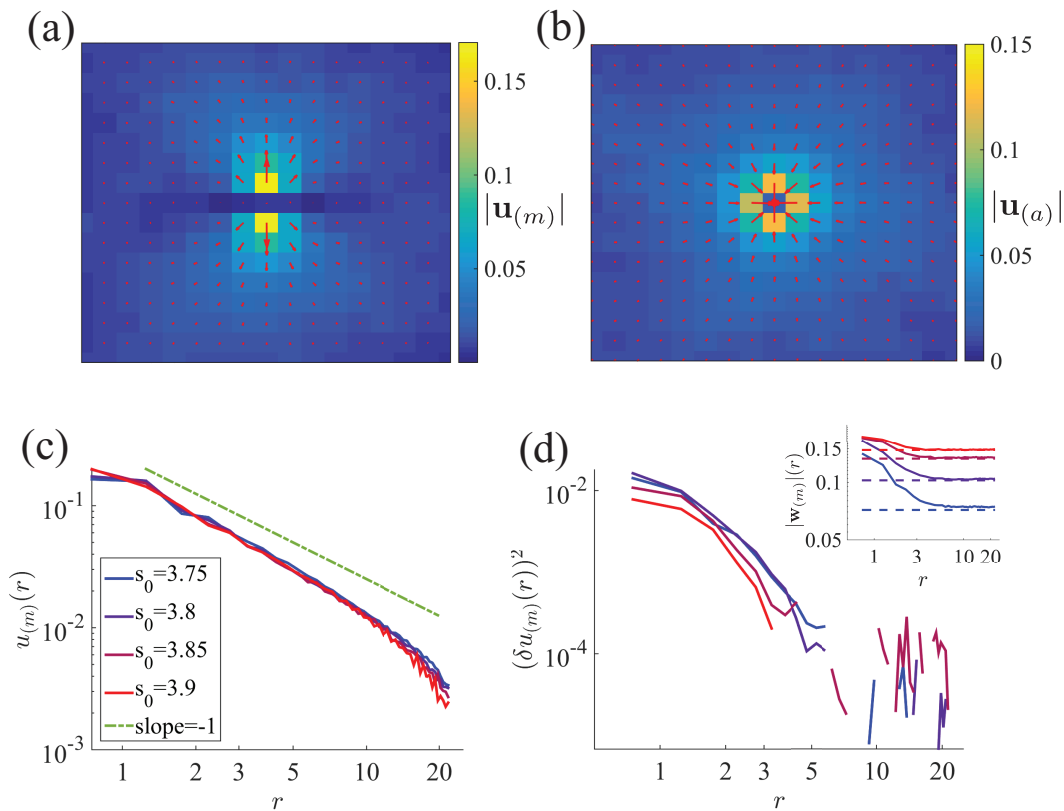}
    \caption{\label{fig:dd_single_events} \mdc{Mean} and \mdc{fluctuating} displacements quantified in response to a single mitosis/apoptosis event. In (a) and (b), arrows show the \mdc{mean} displacements in response to a mitosis and apoptosis, respectively, for a tissue with $s_0=3.6$, $v_0=0.05$. Colormap indicates the magnitude of these \mdc{mean} vectors with grid spacing 1.5. Panels (c) and (d) show cell displacements as a function of distance to the cycling cell for  $s_0 = 3.75, 3.8, 3.85, 3.9$ (blue to red). Panel (c) shows the magnitude of the rotationally averaged \mdc{mean} displacements as a function of distance from a mitosis event. The green dot-dashed line is the expected scaling  $1/r$ . In (d), the values for the $\delta u_{(m,a)}(r)$ per Eq.~\ref{eq:dd_nonaffine_displacements} are plotted similarly, revealing a finite region of \mdc{fluctuating} displacements. Inset shows that $w$ plateaus at long distances to the expected value from mean-square displacement data in the bare AVM.
 }
    \end{center}
\end{figure*}

As noted by Puosi \textit{et. al.}~\cite{Puosi2014} for the similar case of shear transformations in thermal sphere packings, the vector averaged $\mathbf{u}_{(m,a)}(r, \theta)$ captures the \mdc{mean} elastic response of the surrounding medium. %\dms{I feel like rather than go into this discussion, we can just say what we mean.}\lisa{This sentence doesn't make sense to me. Michael, could you re-write to be more specific about it? Note that the term affine is used here for historical reasons, despite being distinct from the mathematical concept of an affine transformation} \mdc{where the Jacobian of the mapping is spatially uniform. For systems which undergo simple shear these two notions of affine do coincide. However, we are considering point-like sources of deformation which do not generally result in a uniform Jacobian. This can be seen in the canonical example of the Eshelby inclusion.}
A straightforward calculation of the response of a homogeneous elastic medium to localized strains~\cite{Puosi2014} suggests that the \mdc{mean} deformations associated with either a cell death or cell division event should fall off as $r^{-1}$ in two dimensions. This is consistent with the numerical observations in our model, shown in Fig.~\ref{fig:dd_single_events}~(a), (b) and (c). In particular, the dashed line in Fig.~\ref{fig:dd_single_events}(c) shows the expected scaling of $r^{-1}$, which is in reasonable agreement with the data. 

 \mdc{Additional cell displacements are expected to arise in the vicinity of the event due to disorder in the tissue structure. These additional \mdc{fluctuating} displacements are captured by $\bm{w}_{(m,a)}(r, \theta)$.} It is important to note that this quantity captures both local contributions from the mitosis (apoptosis) event, as well as contributions everywhere from active, motility driven, cell motions. % This latter portion is expected to dominate at long distances. 
To separate the local contribution of these \mdc{fluctuating} elastic displacements from the noise generated by the active forces in the AVM, we define 
\begin{equation}\label{eq:dd_nonaffine_displacements}
\delta u_{(m,a)}(r)^2 = |\mathbf{w}_{(m,a)}|^2(r) - |\mathbf{w}_{(m,a)}|^2(\infty) \, , 
\end{equation}
where $|w_{(m,a)}|^2(\infty)$ is the far-field plateau in the \mdc{fluctuating} displacement field generated by active noise, shown in the inset to Fig.~\ref{fig:dd_single_events}~(d). The resulting rotationally averaged \mdc{fluctuating} displacement field \mdc{capturing only the motion caused by the cell cycling event}, $\delta u_{(m,a)}(r)$, is plotted in Fig.~\ref{fig:dd_single_events}(d) for a mitosis event.

We note immediately from the data in Fig.~\ref{fig:dd_single_events}(c) that the \mdc{mean} deformations show only minor variation with the change in $s_0$ and therefore depend little on the quiescent dynamics. In contrast, in Fig.~\ref{fig:dd_single_events}(d) we see that the \mdc{fluctuating} motions decrease as we go from more rigid tissue at low $s_0$ to more fluid tissue at $s_0 \sim 3.9$. The degree of variation is, however, still minor and in general the dynamical response to cell division and death events is not expected to be a good indicator of the underlying dynamical state of the tissue.

%%%%%%%%%%%%%%%%%%%%%%%%%%%%%%%%%%%%%%%%%%%%%%%%%%%%%%%%%%%%%%%%%%%%%%%%%%%%%%%%%%%%%%
%%%%%%%%%%%%%%%%%%%%%%%%%%%%%%%%%%%%%%%%%%%%%%%%%%%%%%%%%%%%%%%%%%%%%%%%%%%%%%%%%%%%%%
%%%%%%%%%%%%%%%%%%%%%%%%%%%%%%%%%%%%%%%%%%%%%%%%%%%%%%%%%%%%%%%%%%%%%%%%%%%%%%%%%%%%%%

\subsection{Predicting Cell Displacements from Mitosis and Apoptosis}

%Guided by these single event results, 
%Guided by the displacements measured for single cell-death and cell-division events, we construct simplified models to estimate their influence on measurable quantities, such as $\mathrm{D}$ and $\tau_\alpha$. 
\mdc{In the simple model in Eq.~\ref{eq:dd_taualpha1} it is assumed that motility and cell cycling contribute independently and linearly to the rate of cell motion. This same assumption implies that the diffusion $\mathrm{D}$ will be the sum of independent parts $\mathrm{D} = \mathrm{D}_0 + \mathrm{D}_{\delta}$, where $\mathrm{D}_0$ is the contribution from motile forces and $\mathrm{D}_{\delta}$ is the contribution from the cell cycling. Below, we estimate $\mathrm{D}_{\delta}$ from the displacements measured for single cell-death and cell-division events.}

As pointed out by Ranft \textit{et. al.}~\cite{Ranft2010}, the \mdc{mean} displacements identified above produce a constantly changing reference state in an elastic material. Therefore, while cells may oscillate in their cages, the cages themselves move as a result of each event via the displacement fields $\mathbf{u}_{(m,a)}(r, \theta)$. We would like to estimate the effect of this changing reference state on the mean square displacement of our tracer cells. %as shown by the dashed lines in Fig.~\ref{fig:avm_mitosis_flow}(c),
These events will simultaneously create \mdc{fluctuating} displacements, which will also contribute to the net cell motion. %These events will also be creating \mdc{fluctuating} displacements on top of the \mdc{mean} motions. %\mdc{Quantifying and summing these separate contributions will lead to a prediction for the contribution to diffusion from cell cycling. } %Together, these ingredients should be able to capture the dynamics in the limit of low $\tau_\delta$, where the division and death events should dominate. 

\begin{figure*}[!t]  
\begin{center}
    \includegraphics[width=0.8\textwidth]{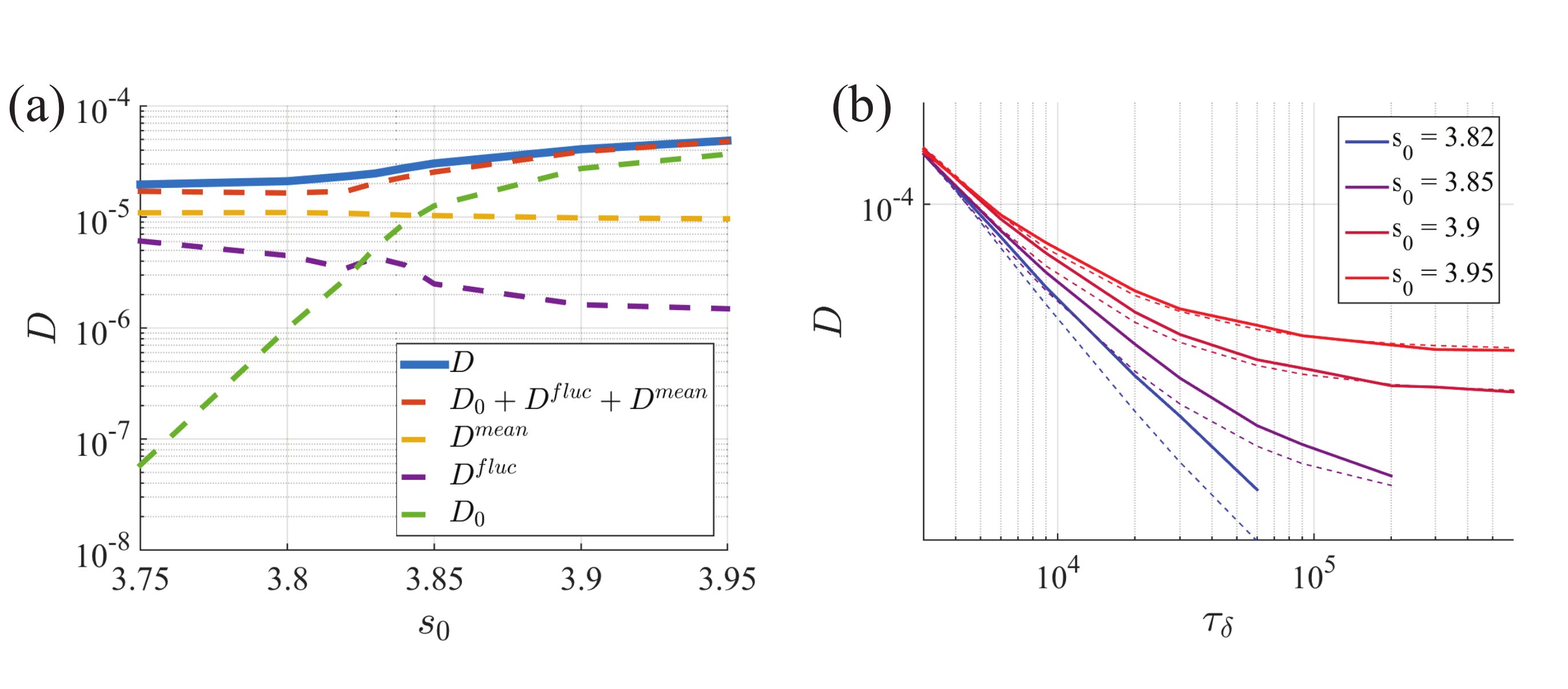}
    \caption{\label{fig:dd_diffusion_check} Separate contributions to diffusion from \mdc{mean} and \mdc{fluctuating} displacements and the bare diffusion add up to approximate the measured diffusion in \textbf{(a)}. In \textbf{(b)}, the validity of this prediction is tested for a range of division rates and a series of $s_0$ values. Simulations where the MSD does not ever become truly diffusive are not included. Criteria for diffusion here is that the MSD-exponent $\alpha$ be greater than the cutoff value $0.9$.}
    \end{center}
\end{figure*}

To estimate the total diffusion produced by these different contributions  (mitosis and apoptosis, \mdc{mean} and \mdc{fluctuating} displacements), we again make the simplifying assumption that they are uncorrelated. We can then write $\mathrm{D}_{\delta}$ as a sum of the decoupled parts
%In practice, the diffusion produced by each of these contributions can be evaluated separately. Assuming these contributions do not display significant correlations (other than in their chosen locations) the diffusion constants from each mechanism may be summed to find the total diffusion due to cell cycling events
%
\begin{equation}\label{eq:dd_total_toy_diffusion}
\mathrm{D}_{\delta} = \mathrm{D}^{mean}_{(a)} + \mathrm{D}^{mean}_{(m)} + \mathrm{D}^{fluc}_{(a)} + \mathrm{D}^{fluc}_{(m)} \, .
\end{equation}
Here, $\mathrm{D}^{mean}_{(m,a)}$ captures the diffusion due to the \mdc{mean} $\mathbf{u}_{(m,a)}(r, \theta)$, while $\mathrm{D}^{fluc}_{(m,a)}$ captures diffusion due to the \mdc{fluctuating} $\delta u_{(m,a)}(r, \theta)$. Quantifying each of these diffusion constants will rely on the simplifying assumptions (1) that we may ignore randomness in the timing of the divisions and deaths and (2) that the effect from each event is felt instantaneously by the surrounding tissue. This enables us to simply sum up the contributions from the $n(t) = t N_{cells} / \tau_\delta$ events that will have taken place after a time $t$. Therefore, each term in Equation~\ref{eq:dd_total_toy_diffusion} may be estimated in terms of an average magnitude of motion per cell per event $d_i$ as
\begin{equation}\label{eq:dd_motion_estimater}
\mathrm{D}_i = \frac{N_{cells}}{4 \tau_\delta} \left\langle d^2\right\rangle_i \, ,
\end{equation}
where the four contributions from the rhs of Eq.~\ref{eq:dd_total_toy_diffusion} are now indexed by $i$. As shown in Appendix~\ref{app:single_events}, the values of $\left\langle d^2 \right\rangle_i$ may be then estimated from summation of the single-event data shown in Fig.~\ref{fig:dd_single_events}(b) and (c). In practice, $\left \langle d^2 \right\rangle_i$ is computed as a spatial average
\begin{equation}\label{eq:dd_displacement_averaging}
\left \langle d^2 \right\rangle_i =\frac{1}{A_T} \int \mathrm{d}^2 \mathbf{x} \, d^2_i(r, \theta)
\end{equation}
where  $d^2_i(r, \theta)$ represents the square displacements captured in our previously averaged values  $\mathbf{u}^2_{(a,m)}(r, \theta)$ and $\delta u^2_{(a,m)}(r, \theta)$. As these simulation data are measured on a radial grid in discrete bins, the integral in Eq.~\ref{eq:dd_displacement_averaging} will turn into a sum over bins, and will have a cutoff radius at $r \sim L/2$. While this excludes cell displacements in the corners of the simulation box, this contribution is small and the predictions constructed without these corners validate their exclusion. As shown in Fig.~\ref{fig:dd_diffusion_check}-(a), each of these contributions to the diffusion $\mathrm{D}_i$ is essential to form a complete prediction of the observed diffusion. In Fig.~\ref{fig:dd_diffusion_check} this prediction provides a reasonable estimate, deviating most significantly for tissues which display glassy signatures in the ($\tau_\delta \rightarrow \infty$ limit) bare-AVM. This breakdown is expected because the corresponding component $\mathrm{D}_0$ in our prediction is no longer truly a diffusion, since the AVM itself displays anomalous subdiffusion in this regime as discussed in Appendix~\ref{sec:AVM}.

This prediction for the diffusion constant may be extended to a prediction for the uncaging time $\tau_\alpha$, which performs similarly well. Relating these two quantities relies on the simple result $ \mathrm{D} \cdot \tau_\alpha = c_B $ which applies to free overdamped Brownian motion in 2D. The constant $c_B \sim 0.1363$ is straightforwardly derived as shown in Appendix.~\ref{app:tau_diff_relation}. Using this relation,  our prediction of $\tau_\alpha$ then relies only on the assumption that, on long time scales, the motion of cells is Brownian.

This is an assumption which, again, begins to break down as motion becomes glassy and cage-escapes become intermittent. In Fig.~\ref{fig:dd_collapses}, the black dashed line prediction matches well with (red) data. However, even where anomalous subdiffusive glassy signatures arise, our prediction captures this crossover between motility-dominated and cell division/death dominated dynamics reasonably well.

\section{Discussion} \label{sec:discussion}
To study the influence of propulsive forces on the dynamical state of tissues, we have developed an Active Vertex Model that naturally generalizes the Vertex Model framework to incorporate cell motility analogously with self-propelled Voronoi models~\cite{Bi2016}. In Section~\ref{sec:AVM}, we have shown that this model qualitatively reproduces some of the dynamical results of the SPV model. As found previously by Bi \textit{et. al.}~\cite{Bi2016}, we locate ``glassy'' states which exhibit sub-diffusive mean square displacements. The transition from a fluid-like state to a solid-like state is achieved either by decreasing the propulsive forces, or by increasing the effective cell-cell interfacial tension.

We then extended the AVM to investigate the influence of cell death and cell division on tissue dynamics. Our investigation was guided by the expectation that the rate of cell division and death will compete with the ``bare'' uncaging rate from the AVM to determine the tissue dynamics. This simple picture allows us to identify evidence of glassy states even at finite rates of apoptosis and mitosis. The glassy behavior found here appears to be in a different regime from the one explored in~\cite{Matoz-Fernandez2017-1}, \mdc{where sources of activity beyond cell cycling were not the focus}. \dms{We again emphasize that another pertinent difference is the choice of whether or not to couple cell cycle events directly to the local structural state of the individual cells.} This may resolve a standing discrepancy between existing experiments and theory. While cell division and death events will generally lead to diffusive fluid-like behavior on the longest timescales, subdiffusive behavior can be observed on intermediate timescales when the bare uncaging rate is slow enough to allow it and the cell division rate is low enough to not obscure it. The subdiffusive dynamics also suggests, by analogy with supercooled fluids, that these tissues will behave elastically on these intermediate timescales.

We further characterized the spatial distribution of cell displacements in the vicinity of each individual mitosis and apoptosis event. As expected from previous work for the case of area-preserving deformations~\cite{Puosi2014}, we have identified an average displacement field that matches the one obtained from elasticity theory, as well as ``\mdc{fluctuating}'' displacements that arise from the disordered geometry. The \mdc{mean} elastic response decays spatially as $1/r$, with a magnitude that does not depend significantly on tissue dynamics. The \mdc{fluctuating} displacements, in contrast,  vary in magnitude with model parameters, such as cell interfacial tension. Specifically, mitosis and apoptosis events tend to generate more rearrangements in rigid tissues than in fluid ones. 

Using the calculated \mdc{mean} and \mdc{fluctuating} displacements resulting from these division and death events, we assess the relative importance of motile forces versus division and death in the tissue dynamics. From this data, we are able to construct predictions for both the diffusion coefficient and the uncaging time. These predictions are very accurate for systems where true diffusive exponents are measured and begin to break down as subdiffusive glassy signatures emerge. Even in these glassy states, however, the prediction gives a good qualitative description of the crossover from motility-dominated to cell cycle-dominated dynamics. 

Taken together, our results provide a general perspective on the role of cell division and cell death in tissues. To reasonably good approximation, cell cycle events appear to produce long-time diffusivity with adds linearly with the existing cell dynamics. This linear behavior appears independent of the approximate cell lifecycle $2 \tau_\delta$ for the values probed here, but is expected to break down as this lifetime gets faster than the speed of sound in the tissue and events begin to interact. 

In addition, we have not addressed the question of whether glassy signatures persist as the tissue size is increased. In Appendix~\ref{subapp:affine_displacements} we present evidence that the total diffusion due to \mdc{mean} displacements scales as $\log(N)$ for our two-dimensional simulations. \mdc{In disordered systems we expect, however, that the signal generated by the \mdc{mean} displacements will destructively interfere with an effective noise generated by motions from other events at a characteristic lengthscale, as has been previously highlighted for sheared particulate packings~\cite{Lemaitre2009}.  % In disordered systems we expect, however, that the signal generated by the \mdc{mean} displacements will be comparable to the noise generated by \mdc{fluctuating} motions at a characteristic lengthscale, as has been previously highlighted for sheared particulate packings~\cite{Lemaitre2009}.
Beyond this lengthscale, we no longer expect the $\log(N)$ scaling to hold, and so the system will remain glassy in the thermodynamic limit. While we do not identify such a lengthscale for the simulations performed here, testing this hypothesis is an important avenue for future investigations.} %Testing this hypothesis carefully is therefore an important avenue for future investigations.

\lisa{So far, we have focused on dynamics that explicitly keeps the tissue at a fixed number density, in order to pinpoint the effect of cell divisions on tissue dynamics. However, in many experiments on epithelial monolayers, a common and interesting regime is where cell divisions outnumber cell deaths and so the number density increases rapidly. An important open question is how Active Vertex Model parameters, such as $s_0$ or $v_0$, change during epithelial densification. Several experiments on pairs or triplets of cells indicate that many epithelial cells exhibit ``contact inhibition of locomotion'', where cells reduce traction forces and slow down upon contact~\cite{Puliafito2012, Abercrombie1970}. Such behavior could be modeled by decreasing $v_0$ as a function of time in densifying monolayers, and additional experiments to measure traction forces and fluctuations during such processes would be very valuable in constraining model predictions.  Perhaps even more interestingly, it remains unclear how cells regulate their shape $s_0$ during such processes.  One possibility is that cells attempt to maintain the same shape despite changes to density, meaning that they must change their perimeter to match their decreasing area as a function of time.  Alternatively, one could postulate that cells keep their preferred perimeter fixed while their area decreases dramatically, resulting in an increasing $s_0$ that would fluidize a tissue. This latter hypothesis seems inconsistent with existing experimental data, where tissues are generically found to solidify as they become more dense.  Careful studies of cell shape coupled with careful studies of interfacial tensions~\cite{Saraswathibhatla2019} would help to constrain models and better test model predictions. 

Finally, it is interesting to think about how cell division and death rates are themselves affected by tissue dynamics.  For example, it is known that in many cell types cell division rates are governed by the magnitude of local stresses that build up in the tissue~\cite{Shraiman2005, Streichan2014}. In addition, the \mdc{orientation} of a cell division is also controlled by local stresses. This creates the possibility for interesting feedbacks, where tissue dynamics is controlled by the rate of cell division and the rate of cell division is controlled by tissue dynamics.  Understanding precisely how division and death affect dynamics is therefore very important for predicting how such feedback loops can control tissue growth and patterning. It would be interesting to explore how such feedback loops generate patterns in AVMs, and compare to experiment.}

%\mdc{Finally, our results are only suggestive of rheological measures in tissues of constant density. However, real tissues vary their number density dramatically during the process of epithelialization. If we imagine that none of our Active Vertex Model parameters actually change considerably with density, then it would seem that increasing density will increase the effective value of the tuning parameter $s_0$ and fluidize the tissue. The opposite phenomenon is ubiquitously found in experiments on real tissues, which rigidify and slow as the number of cells grows to a steady state value. The standard explanation of this observation has thus far come from "Contact Inhibition of Locomotion", whereby cells sense the changing number density and effectively stop producing crawling forces. However, it is entirely possible that other mechanical parameters change and contribute to the dynamic process of tuning across this supercooling transition.
%Careful modeling and simulation may prove essential in parsing different possible causes of this transition in living epithelial tissues.} 

-%dom lbeinear inatedes our searchstufy i for a scaling collapse of rheological data across a range of parameters. While the results suggest a universal behavior in the crossover from tissues which are effectively unaffected by cell division \& death to tissues where these events will dominate, the data do not fully collapse onto a single curve. The remaining spread in the data may be indicative of a few different effects: (1) that the tissue structural relaxation rate cannot be expressed simply as a sum of relaxation rates from separate sources of activity or (2) that cell division and death produce a structural change in a tissue, altering the role of motile forces in fluidizing the tissue or (3) that the displacements produced by cell division and by cell death events are not independent, but interfere and correlate with one another in time and space. This third possibility is supported by diffusion data, where the whole is not found to be the sum of the separate contributions from the mitosis, apoptosis, and motility. In our future investigations, we will attempt to further parse and quantify these effects.

%Finally, we investigate the scaling of the $\tau_\alpha$ with increasing the system size. A scaling collapse of this data is guided by a toy model for the affine displacements, and reveals that the tissue diffusion increases as $Log(N_{cells})$. This seems to imply that the glassy signatures identified herein will be washed out in the thermodynamic limit. However, for very large systems more cell cycling events will be happening simultaneously and are expected to interfere destructively. It therefore remains unclear what rheological signatures will be observed in the large system limit, such investigations are reserved for future work and may be better addresses with continuum analytic tools.

\paragraph*{Acknowledgements}
This work was supported primarily by Simons Foundation Grants Number 446222 and 342354 (MC and MCM). Partial support also came from: NSF-DGE-1068780 (MC), NSF-DMR-1609208 (MCM and MC), NSF-DMR-1352184 (MC and MLM), NSF-PHY-1607416 (MLM and DMS) as well as Simons Foundation Grant Number 454947 (MLM and DMS).

%%%%%%%%%%%%%%%%%%%%%%%%%%%%%%%%%%%%%%%%%%%%%%%%%%%%%%%%%%%%%%%%%%%%%%%%%%%%%%%%%%%%%%
%%%%%%%%%%%%%%%%%%%%%%%%%%%%%%%%%%%%%%%%%%%%%%%%%%%%%%%%%%%%%%%%%%%%%%%%%%%%%%%%%%%%%%
%%%%%%%%%%%%%%%%%%%%%%%%%%%%%%%%%%%%%%%%%%%%%%%%%%%%%%%%%%%%%%%%%%%%%%%%%%%%%%%%%%%%%%

\section{Appendices}
\appendix
\section{The Active Vertex Model}

Here we include a more detailed description of the Active Vertex Model implemented in this work.

\subsection{Forces in the Active Vertex Model\label{app:avm_explained}}
In this Active Vertex Model (AVM), the degrees of freedom are the positions of the vertices of the spatial tiling. The vertex labeled $\mu$ follows an overdamped equation of motion
\begin{equation}\label{eq:app_vert_dynamics}
\partial_t \mathbf{r}^\mu = -\frac{1}{\gamma} \frac{\partial}{\partial\mathbf{r}^\mu} E \, ,
\end{equation}
where $\gamma$ is the substrate friction. The effective energy
\begin{equation}\label{eq:app_total_energy}
E = E_{shape}  - \gamma v_0 \sum_{cells-a}\mathbf{\hat{n}}(\theta_a) \cdot \mathbf{r}_a\, ,
\end{equation}
captures the self-propulsion force (of magnitude $v_0$) of each cell $a$ in direction $\theta_a$ in addition to the standard shape energy terms from Eq.~\ref{eq:AVM_evertex}. The geometric center (centroid) $\mathbf{r}^a$ of cell $a$ defined by
\begin{equation}\label{eq:AVM_centroid}
\mathbf{r}_a = \frac{1}{6 A_a} \sum_{\nu} \left( \mathbf{r}_a^\nu  + \mathbf{r}_a^{\nu+1} \right) (\mathbf{r}_a^\nu  \times \mathbf{r}_a^{\nu+1}) \cdot \mathbf{\hat{k}} \, ,
\end{equation}
captures the center of mass of a polygon of uniform mass density. Here, $\nu$ indexes the $N_a$ vertices on cell $a$ in counterclockwise fashion and $\mathbf{r}_a^{N_a + 1} = \mathbf{r}_a^{1}$. The polygon area may be expressed in similar terms as
\begin{equation}\label{eq:AVM_area_formula}
A_a = \frac{1}{2} \sum_{\nu} (\mathbf{r}_a^\nu  \times \mathbf{r}_a^{\nu+1}) \cdot \mathbf{\hat{k}} \, .
\end{equation}

\begin{figure}[!t]  
\begin{center}
    \includegraphics[width=0.48\textwidth]{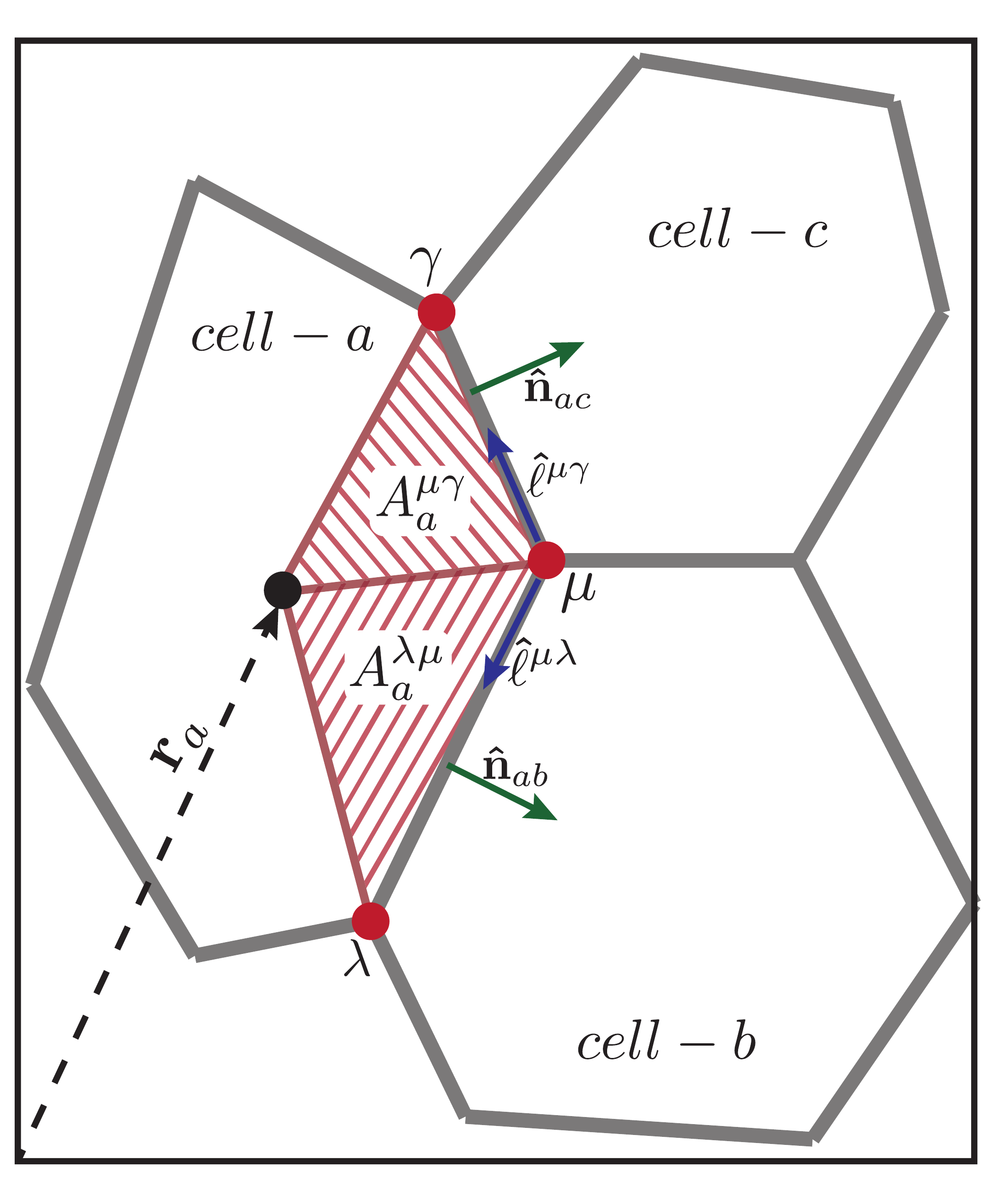}
    \caption{\label{fig:app_shape_forces}Illustration of the vectors and geometric components involved in evaluating the forces on vertex $\mu$ due to the shape energy and the motility of cell $a$}
    \end{center}
\end{figure}

The force (and therefore motion) on each vertex can be calculated by carrying out the derivatives in Eq.~\ref{eq:AVM_vert_dynamics}. Each vertex in this model is connected to 3 cells and each cell energy will contribute separate terms to the net motion of the vertex. If vertex $\mu$ is connected to cells $a$, $b$ and $c$, then the motion breaks down into
\begin{equation}
\begin{aligned}\label{eq:AVM_vert_dynamics2}
\partial_t \mathbf{r}^\mu& = -\frac{1}{\gamma} \left( \frac{\partial}{\partial\mathbf{r}^\mu} E_a + \frac{\partial}{\partial\mathbf{r}^\mu} E_b + \frac{\partial}{\partial\mathbf{r}^\mu} E_c  \right) \\ & + v_0 \left( \frac{\partial (\mathbf{\hat{n}}(\theta_a) \cdot \mathbf{r}_a) }{\partial \mathbf{r}^\mu} + \frac{\partial (\mathbf{\hat{n}}(\theta_b) \cdot \mathbf{r}_b)}{\partial \mathbf{r}^\mu} + \frac{\partial (\mathbf{\hat{n}}(\theta_c) \cdot \mathbf{r}_c )}{\partial \mathbf{r}^\mu} \right)\, .
\end{aligned}
\end{equation}
For simplicity, we may focus on the contributions from cell $a$. As identified in previous work~\cite{Yang2017}, the shape energy produces tension-based and pressure-based forces on each vertex. The shape-based force on vertex $\mu$ from cell $a$ reads
\begin{equation}\label{eq:AVM_vertex_shape_force}
\frac{\partial}{\partial\mathbf{r}^\mu} E_a = -\frac{\Pi_a}{2} (\mathbf{\hat{n}}_{ab} l^{\mu\gamma} + \mathbf{\hat{n}}_{ac} l^{\mu\lambda}) -T_a (\mathbf{\hat{l}}^{\mu\gamma} + \mathbf{\hat{l}}^{\mu\lambda}) \, ,
\end{equation}
where $\gamma$ and $\lambda$ index the vertices of cell $a$ which are adjacent to $\mu$, $l^{\mu\gamma}$ ($l^{\mu\lambda}$) and $\mathbf{\hat{l}}^{\mu\gamma}$ ($\mathbf{\hat{l}}^{\mu\lambda}$) are the length and direction of the edge connecting vertex $\mu$ to vertex $\gamma$ ($\lambda$) and the unit vector $\mathbf{\hat{n}}_{ab}$ ($\mathbf{\hat{n}}_{ac}$) points across the edge shared by cell $a$ and cell $b$ ($c$) as in Fig~\ref{fig:app_shape_forces}. 
The tension and pressure of cell $a$ are respectively
\begin{equation}\label{eq:AVM_pressure_tension}
T_a = \frac{\partial E_{shape}}{\partial P_a} \qquad \qquad \Pi_a = - \frac{\partial E_{shape}}{\partial A_a}\, .
\end{equation}

Similar to the above, we now evaluate derivatives to understand the self propulsion forces which act on the vertices. Again, we will consider only the contributions from cell $a$. The derivative will take the form
\begin{equation}\label{eq:AVM_dcentroid1}
\frac{\partial (\mathbf{\hat{n}}(\theta_a) \cdot \mathbf{r}_a) }{\partial \mathbf{r}^\mu} = \frac{\partial  \mathbf{r}_a }{\partial \mathbf{r}^\mu} \cdot \mathbf{\hat{n}}(\theta_a)
\end{equation}
With these expressions, taking the derivative in Eq.~\ref{eq:AVM_dcentroid1} is tedious but straightforward. This becomes
\begin{equation}
\begin{aligned}\label{eq:AVM_dcentroid2}
\frac{\partial r_{a,j}}{\partial r^\mu_{i}} &= \frac{1}{6 A_a} \Bigg[ \delta_{ij} (\mathbf{r}^\lambda \times \mathbf{r}^\mu)\cdot\mathbf{\hat{k}} + \delta_{ij} (\mathbf{r}^\mu \times \mathbf{r}^\gamma)\cdot\mathbf{\hat{k}} \\ & + (r_j^\lambda + r_j^\mu) R(\pi/2)_{ik} r^\lambda_k + (r_j^\mu + r_j^\gamma) R(-\pi/2)_{ik} r^\gamma_k \\ & -3 (R(\pi/2)_{ik} r^\lambda_k + R(-\pi/2)_{ik} r^\gamma_k) r_{a,j} \Bigg] \, ,
\end{aligned}
\end{equation}
where
\begin{equation}\label{eq:AVM_rotation}
R(\theta)_{ij} = 
\left[
  \begin{tabular}{cc}
  $\cos(\theta)$ &  $-\sin(\theta)$ \\
  $\sin(\theta)$ &  $\cos(\theta)$
  \end{tabular}
\right]
\, 
\end{equation}
is a vector rotation by an angle $\theta$, $(i,j)$ index cartesian vector components in the $x-y$ plane and Einstein summation convention is assumed for repeated indices. While the expression in Eq.~\ref{eq:AVM_dcentroid2} is a bit unwieldy, we note that it is translationally invariant and may consider it in a more convenient coordinate system $\tilde{\mathbf{r}} = \mathbf{r} - \mathbf{r}_a$ which has its origin at $\mathbf{r}_a$. Using this, we find the force on vertex $\mu$ due to the motility of cell $a$ as
\begin{equation}
\begin{aligned}\label{eq:AVM_motile_vforce}
\mathbf{f}^{\mu(a)}_{i} &= v_0 \hat{n}(\theta_a)_{j}\frac{\partial r_{a,j}}{\partial r^\mu_{i}} \\ & = \frac{v_0 \hat{n}(\theta_a)_i}{3 A_a} [ A_a^{\lambda\mu} + A_a^{\mu\gamma} ] \\ & + \frac{v_0}{6 A_a} \Bigg[ R(\pi/2)_{ik} \tilde{r}^\lambda_k \hat{n}(\theta_a)_{j} (\tilde{r}_j^\lambda + \tilde{r}_j^\mu) \\ &  + R(-\pi/2)_{ik} \tilde{r}^\gamma_k  \hat{n}(\theta_a)_{j} (\tilde{r}_j^\mu + \tilde{r}_j^\gamma)   \Bigg] \, ,
\end{aligned}
\end{equation}
where $ A_a^{\lambda\mu}  $ and $ A_a^{\mu\gamma} $ are the area of triangles with vertices $\{\mathbf{r}_a,\mathbf{r}^\lambda, \mathbf{r}^\mu  \}$ and $\{\mathbf{r}_a,\mathbf{r}^\mu, \mathbf{r}^\gamma  \}$ respectively, as shown in Fig.~\ref{fig:app_shape_forces}.
While this expression does not lend itself to much insight, we can see on inspection that the first terms will move the vertex in the direction of $\mathbf{\hat{n}}(\theta_a)$. The second terms in the large square brackets will roughly serve to shrink the length of interfaces at the back of the cell, while expanding the length of interfaces at the front. 

\begin{figure}[!t]  
\begin{center}
    \includegraphics[width=0.48\textwidth]{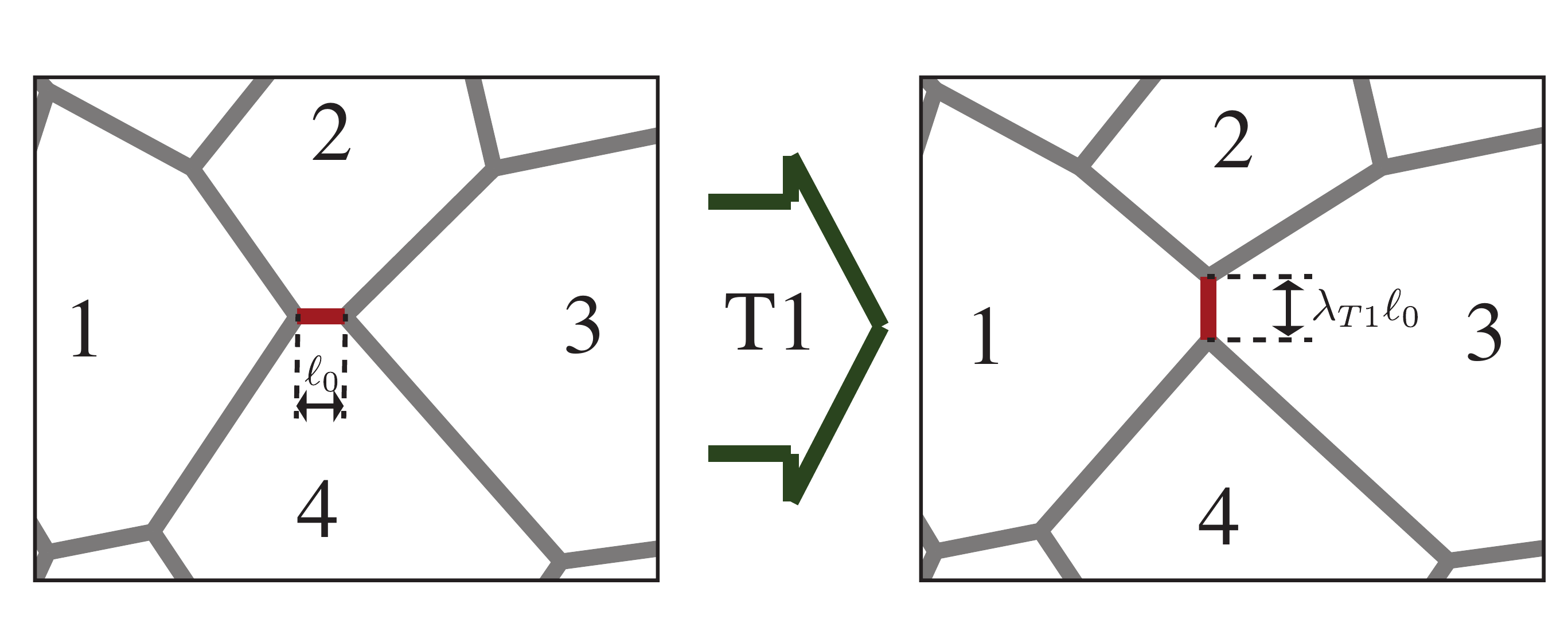}
    \caption{\label{fig:t1}A T1 topological rearrangement in our AVM simulations identifies an edge with length $l_0 < l_c$,  rewires the network connections appropriately, rotates by $\pi/2$, and extends the edge length to a factor $\lambda$ times its original length.}
    \end{center}
\end{figure}

While these forces determine the continuous time-evolution of the vertices, there can be no large-scale migration of cells through the tissue until topological rearrangements are allowed. We therefore include a protocol for T1 rearrangements (as shown in Fig~\ref{fig:t1}) which are sufficient to explore the space of cellular topological configurations at constant density. In practice, the edge lengths are periodically checked after a time $t_{T1}$ for values lower than a threshold $\ell_c$. These edges are then topologically rearranged so that the two cells which initially share the short edge are no longer in contact. The edge is then rotated by $\pi/2$ and the length extended by a factor $\lambda_{T1}$. The value of $t_{T1}=0.05$ used here is chosen for speed of simulation, while the rescaling factor $\lambda_{T1}=2$ is chosen to avoid ``T1-traps'' whereby the same transition may repeat itself regardless of energetic favorability. The cutoff length $\ell_c = 0.04$ is chosen small enough to make the transition appear continuous but also large enough so that a vertex may ``find'' and activate the desired T1. The parameters $\ell_c$, $\lambda_{T1}$ and $ t_{T1}$ have been separately varied within reasonable ranges and the impact on the dynamics appears insubstantial. These simulations are implemented using the ``cellGPU'' codebase previously developed by one of us~\cite{Sussman2017-1}.

We note again that this form of the Active Vertex Model is chosen to minimize the differences with the Self-Propelled Voronoi model. The differences are limited to the following: (1) the Active Vertex Model has more degrees of freedom, avoiding the shape constraints of Voronoi tesselations, (2) T1 rearrangements in the AVM must be done by hand, while in the SPV they come about naturally and (3) motility in the AVM is designed to propel the centroid of the cell, whereas for the SPV this is replaced by the Voronoi center.

%%%%%%%%%%%%%%%%%%%%%%%%%%%%%%%%%%%%%%%%%%%%%%%%%%%%%%%%%%%%%%%%%%%%%%%%%%%%%%%%%%%%%%
%%%%%%%%%%%%%%%%%%%%%%%%%%%%%%%%%%%%%%%%%%%%%%%%%%%%%%%%%%%%%%%%%%%%%%%%%%%%%%%%%%%%%%
%%%%%%%%%%%%%%%%%%%%%%%%%%%%%%%%%%%%%%%%%%%%%%%%%%%%%%%%%%%%%%%%%%%%%%%%%%%%%%%%%%%%%%

\subsection{Subdiffusive Exponents and Exotic Dynamical States in the Active Vertex Model \label{app:avm_rheology}}

%As noted in the main text, we now probe the rheological signatures of glassiness in our Active Vertex Model. This is accomplished using the mean-square displacement of cells and using the Overlap as defined in Eq.~\ref{eq:overlap_definition}. 
While the main text describes the glassy behavior present in the AVM, Vertex Models governed by an energy of the type in Eq.~\ref{eq:AVM_evertex} may also be subject to anomalous energy landscapes and non-trivial inherent stresses~\cite{Sussman2017-2, Sussman2018-2}. As a reflection of these features, dynamical signatures such as the mean-square displacement (MSD) and the self-overlap (as defined in Eq.~\ref{eq:AVM_evertex}) may appear anomalous in the vicinity of the glassy phase. Here, we explore these anomalies in greater detail.  

The mean-square displacement in thermal particulate systems approaching a glass transition will display the signatures of ``caging''. This manifests as a flat plateau in the MSD as particles reach this cage size (radius) and are temporarily trapped before diffusing away on a longer timescale~\cite{Berthier2011, Weeks2002}. \mdc{The AVM MSD, in contrast, near the onset of glassiness does not appear to completely flatten at any discernible length scale}. Instead, we observe an extended subdiffusive regime which emerges as the system approaches the glassy state, implying that there may not be a well defined cage-size and that the cells are exploring an unusual distribution of local metastable states as they attempt to realize their target shapes. It may therefore not be useful to think of the travel between these metastable states as "cage-breaking" which come paired with intermittent jumps in displacement on the order of the particle size. Instead, it appears that a distribution of effective cages may be traversed by the cells in a more continuous manner. 

\begin{figure}[!t]  
\begin{center}
    \includegraphics[width=0.48\textwidth]{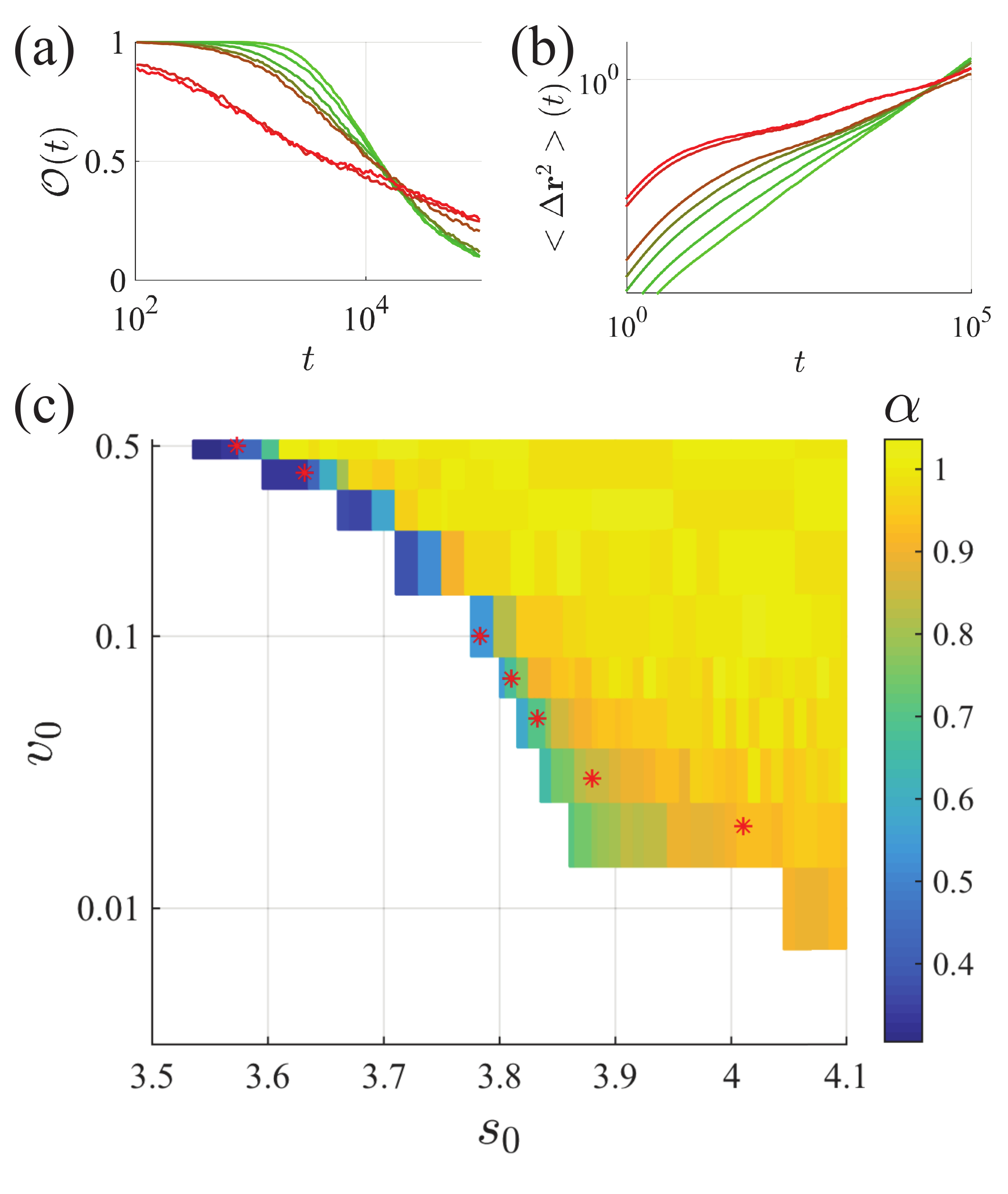}
    \caption{\label{fig:app_avm_subdiffusive} Revealing the persistent subdiffusion in the Active Vertex Model. In \textbf{c}, the long-time exponent of the MSD is plotted on a colormap for the same phase data as in main text Fig.~\ref{fig:AVM_phase_behavior}c. Phase points where the uncaging time $\tau_{\alpha0}$ is beyond the simulation time are left white. Along the interface, a region where this exponent $\alpha$ is distinctly lower than unity despite the finite uncaging time indicates the anomalous dynamics. In \textbf{a}, and \textbf{b}, we display the varied character of the overlap \textbf{(a)} and the MSD \textbf{(b)} curves for points with the same uncaging time $\tau_{\alpha0} \sim 2 \cdot 10^4$ and corresponding, from green to red, to the points at $(s_0, v_0) =$ (4.01, 0.02), (3.88, 0.03), (3.833, 0.05) , (3.81, 0.07), (3.783, 0.1), (3.632, 0.4), (3.573, 0.5) as indicated by the red stars in the phase diagram. All data for $D_r = 1.0$ and $N_{cells} = 300$. }
    \end{center}
  \end{figure}

\begin{figure*}[!t]  
\begin{center}
    \includegraphics[width=0.98\textwidth]{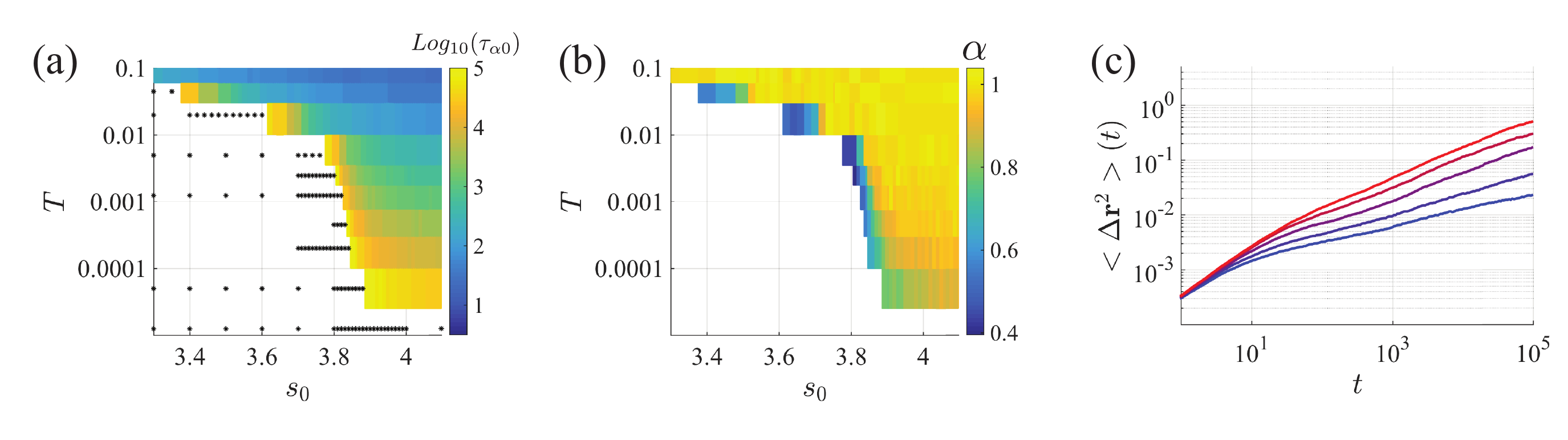}
    \caption{\label{fig:app_avm_thermal} Summarizing the dynamical signatures in the Thermal Vertex Model. In \textbf{a}, the uncaging time $\tau_{\alpha0}$ is displayed in a logarithmic colormap revealing similar behavior to that for the AVM shown in Fig.~\ref{fig:AVM_phase_behavior}c. Black asterisks again indicate phase points where $\tau_{\alpha0}$ is out of range. In \textbf{b}, the long-time MSD exponent $\alpha$ is plotted in a colormap for the same phase space in \textbf{a}, again with the points where the uncaging time is out of range left blank. A region of phase space with subdiffusive behavior extending sufficiently beyond cage escape is indicated by the dark region near the boundary. In \textbf{c}, the extended subdiffusion is shown for points (from blue to red) at $s_0 = (3.75, 3.77, 3.79, 3.81, 3.83)$ and $T=0.00045$. All data for $N_{cells} = 300$. }
    \end{center}
  \end{figure*}

The above points to the unconventional geometric mechanics of the Vertex Model as the cause of this anomalous behavior, but we must also consider the details of the non-equilibrium dynamics chosen here. As we have employed a non-zero persistence time of these forces as $\tau_p = 1/\mathrm{D}_r = 1$, we also perform simulations with the motility replaced by thermal forces on the vertices at effective temperature $T$. \mdc{These thermal} simulations are run in the AVM with $v_0$ set to zero, and the equation of motion of a vertex instead obeying
\begin{equation}\label{eq:app_thermalvertex}
\partial_t \mathbf{r}^\mu = -\frac{1}{\gamma} \frac{\partial}{\partial\mathbf{r}^\mu} E + \mathbf{\eta}^{\mu} \,   
\end{equation}
where
\begin{equation}\label{eq:app_brownian_average}
<\mathbf{\eta}^{\mu}(t)> = 0\, 
\end{equation}
and
\begin{equation}\label{eq:app_brownian_variance}
<\eta_i(t_1)\eta_j(t_2)> = \frac{2 T}{\gamma} \delta_{ij} \delta(t_1 - t_2) \, 
\end{equation}
Results for this thermal vertex model are summarized in Fig.~\ref{fig:app_avm_thermal} and indicate qualitatively the same behavior as observed for cell motile forces. We also note that previous studies in the Self-Propelled Voronoi and Thermal Voronoi models have not explicitly identified this subdiffusive dynamical feature. This indicates that this strange extended subdiffusion is associated with the Vertex Model geometric energy along with the vertex degrees of freedom that it acts on.

\begin{figure}[!t]  
\begin{center}
    \includegraphics[width=0.48\textwidth]{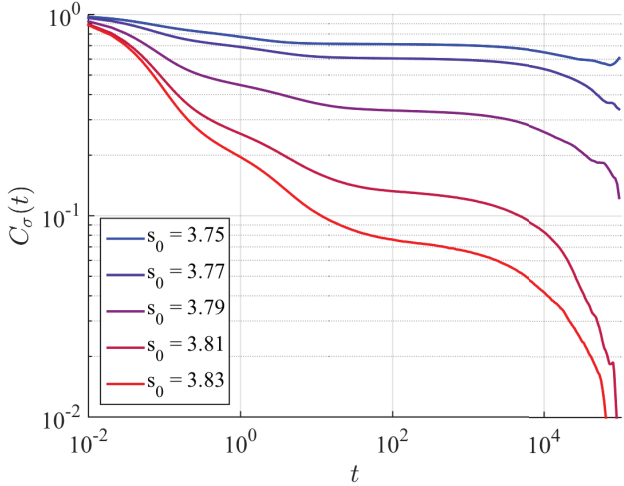}
    \caption{\label{fig:app_thermal_stresscorr} The shear-stress autocorrelation function $C_{\sigma}(t)$ in the Thermal Vertex Model shows unusual temporal character in the glassy state. This data for $T=0.00045$ over a series of $s_0$ values at the same points as in Fig.\ref{fig:app_avm_thermal}c. Three timescales are suggested instead of the traditional two ``alpha'' and ``beta'' relaxation times.  }
    \end{center}
  \end{figure}

Finally, we investigate temporal fluctuations in the mechanical stress as defined instantaneously by
\begin{equation}\label{eq:mech_stress}
  \sigma_{ij} = \frac{1}{A_T} \sum_{cells-a} A_a \left[ \frac{\partial E_{shape}}{\partial A_a} \delta_{ij} + \frac{1}{2 A_a} \sum_{\mu\nu \in a}\frac{\partial E_{shape}}{\partial l^{\mu\nu}_i} \right]
\end{equation}
as in Ref.~\cite{Yang2017}. From this global tissue stress, we compute the stress autocorrelation function via
\begin{equation}\label{eq:app_stresscorr}
  C_{\sigma}(t) = \frac{ \left\langle \sigma_{xy}(t_0 + t) \sigma_{xy}(t_0) \right\rangle_{t_0} }{ \left\langle \sigma_{xy}^2(t_0) \right\rangle_{t_0}} \, .
\end{equation}
Where the angular brackets denote the average over both noise and over the time $t_0$. In practice, the computation of this correlation is accomplished using the multiple-time algorithm described by Ram{\'i}rez \textit{et. al.} in Ref.~\cite{Ramirez2010}. This stress auto-correlation function, shown in Fig.~\ref{fig:app_avm_thermal} for a range of $s_0$ values, indicates additional unusual behavior. Specifically, we do not resolve the traditionally expected ``beta'' (fast) and ``alpha''(slow) decorrelation times. Instead, the stress appears to decorrelate in three or more stages across a range of timescales. This may also indicate that the process of cage breaking can occur across a broad range of different length scales (and therefore different time scales) in epithelial tissues.

%%%%%%%%%%%%%%%%%%%%%%%%%%%%%%%%%%%%%%%%%%%%%%%%%%%%%%%%%%%%%%%%%%%%%%%%%%%%%%%%%%%%%%
%%%%%%%%%%%%%%%%%%%%%%%%%%%%%%%%%%%%%%%%%%%%%%%%%%%%%%%%%%%%%%%%%%%%%%%%%%%%%%%%%%%%%%
%%%%%%%%%%%%%%%%%%%%%%%%%%%%%%%%%%%%%%%%%%%%%%%%%%%%%%%%%%%%%%%%%%%%%%%%%%%%%%%%%%%%%%

\section{Relating the Uncaging Time to the Diffusion Constant for Brownian Particles\label{app:tau_diff_relation}}

This work relies largely on two scalar measures of tissue dynamics, the uncaging time $\tau_\alpha$ and the diffusion coefficient $\mathrm{D}$. Here we show that, for the case of Brownian motion in two dimensions, these quantities may be related explicitly as
\begin{equation}\label{eq:app_tau_diff_relation}
\tau_\alpha \mathrm{D} = c_B \equiv \frac{- b^2}{ 4 \ln[1 - \frac{1}{e}]} \, ,
\end{equation}
where $b$ is the radius defining the cage size in the overlap function (Eq.~\ref{eq:overlap_definition}). This equation (which is related to the Stokes-Einstein relation) breaks down as a material approaches a glass transition, and therefore provides important context for interpreting dynamical parameters. 

To understand Eq.~\ref{eq:app_tau_diff_relation}, we note that free Brownian motion of an ensemble of particles may be described by a smooth probability density function $P(\mathbf{x}, t)$ which is governed by Fick's Law
\begin{equation}\label{eq:app_fick}
\partial_t P(\mathbf{x}, t) = \mathrm{D} \nabla^2 P(\mathbf{x}, t) \, .
\end{equation}
To understand the measured uncaging time, we must first solve for the $P(\mathbf{x}, t)$ for a single particle starting at an arbitrary point $\mathbf{x}_0$ at a time $t_0$. This corresponds to solving with an initial condition $P(\mathbf{x}_0, t_0) = \delta(\mathbf{x} - \mathbf{x}_0)$. Preserving the normalization of $P$, this is solved for $t>t_0$ by
\begin{equation}\label{eq:app_pdf_brownian}
P(\mathbf{x}, t) = \frac{1}{4 \pi \mathrm{D} (t-t_0)} \textrm{Exp}\left\{\frac{-\left[\mathbf{x} - \mathbf{x}_0\right]^2}{4 \mathrm{D} (t-t_0)}\right\} \, .
\end{equation}
This probability distribution may be used to take averages and we may therefore rewrite the self-overlap,
\begin{equation}\label{eq:app_alt_overlap}
\mathcal{O}(t) = \left\langle \Theta( b - |\Delta \mathbf{r}_i(t)|) \right\rangle \, ,
\end{equation}
as 
\begin{equation}\label{eq:app_alt_overlap2}
\mathcal{O}(t) = \int_0^b r \mathrm{d}r \int_0^{2 \pi} \mathrm{d}\theta P(r, t)  \, ,
\end{equation}
where, for convenience, we have set $t_0 =0$ and $r = |\mathbf{x} - \mathbf{x}_0|$. The uncaging time is then defined by the condition $\mathcal{O}(\tau_\alpha) = 1/e$ and we may insert Eq.~\ref{eq:app_pdf_brownian} into Eq.~\ref{eq:app_alt_overlap2}, integrate, and manipulate to find our desired result in Eq.~\ref{eq:app_tau_diff_relation}. Evaluating this constant for the case $b=0.5$, we find that $\tau_\alpha \mathrm{D} = c_B \sim 0.1363$.

%%%%%%%%%%%%%%%%%%%%%%%%%%%%%%%%%%%%%%%%%%%%%%%%%%%%%%%%%%%%%%%%%%%%%%%%%%%%%%%%%%%%%%
%%%%%%%%%%%%%%%%%%%%%%%%%%%%%%%%%%%%%%%%%%%%%%%%%%%%%%%%%%%%%%%%%%%%%%%%%%%%%%%%%%%%%%
%%%%%%%%%%%%%%%%%%%%%%%%%%%%%%%%%%%%%%%%%%%%%%%%%%%%%%%%%%%%%%%%%%%%%%%%%%%%%%%%%%%%%%

\section{Displacements due to single division or death events\label{app:single_events}}

Here we describe our method of converting displacement data for single apoptosis and mitosis events into averaged motions per event $ \left\langle d^2 \right\rangle_i$ as well as a predicted diffusion coefficient $\mathrm{D}_\delta$ per Equations~\ref{eq:dd_total_toy_diffusion} and \ref{eq:dd_displacement_averaging}.

%%%%%%%%%%%%%%%%%%%%%%%%%%%%%%%%%%%%%%%%%%%%%%%%%%%%%%%%%%%%%%%%%%%%%%%%%%%%%%%%%%%%%%
\subsection{Estimating \mdc{mean} contributions to displacement}\label{subapp:affine_displacements}
%To estimate the first and second terms on the right hand side of Eq.~\ref{eq:dd_total_toy_diffusion}, we consider the displacement of a tracer cell as a result of the affine motion $\mathbf{u}_{(m,a)i}(r, \theta)$ produced in each cell cycle $i$. With the dynamics
% measure the corresponding average-squared affine displacement per mitosis or apoptosis event $ \left\langle d^2 \right\rangle^{aff}_{(m,a)}$. This is accomplished
To estimate the first and second diffusion coefficients on the right hand side of Eq.~\ref{eq:dd_total_toy_diffusion}, we will use the "single-event" \mdc{mean} motion data $\mathbf{u}_{(m,a)i}(r, \theta)$ obtained in Section~\ref{sec:dd_single_events}. On the long timescales which we are interested in, each event effectively instantaneously moves the cells via the \mdc{mean} displacement $\mathbf{u}_{(m,a)i}(r, \theta)$ and the cells then follow an equation of motion
\begin{equation}\label{eq:app_affine_dynamics}
\Delta \mathbf{x}_{(m,a)}(t) = \sum_{i}^{n(t)} \mathbf{u}_{(m,a)}(r_i, \theta_i) \, 
\end{equation}
in which the $n(t)$ displacements due to previous events are simply summed up. 
With this EOM, the MSD may be computed as
\begin{equation}\label{eq:app_affine_msd}
\langle\Delta \mathbf{x}^2_{(m,a)}(t)\rangle = \sum_{i}^{n(t)} \left \langle d^2 \right\rangle^{mean}_{(m,a)} \, .
\end{equation}
where we have implicitly defined the average \mdc{mean} motion per event $\left \langle d^2 \right\rangle^{mean}_{(m,a)}$ and $\langle \rangle$ represents the average over the realizations of these apoptosis and mitosis events. As the stochasticity here comes from the spatial positioning (and orientation) of the mitosis (apoptosis) event with respect to our tracer, finding this average for $ \left \langle d^2 \right\rangle^{mean}_{(m,a)} = \langle\mathbf{u}^2_{(m,a)i}\rangle $ requires integrating over the possible positions and orientations of this mitosis (apoptosis) event. This amounts to a spatial integral
\begin{equation}\label{eq:app_displacement_averaging_affine}
\left \langle d^2 \right\rangle^{mean}_{(m,a)} =\frac{\rho}{N_{cells}} \int \mathrm{d}^2 \mathbf{x} \, \mathbf{u}^2_{(m,a)i}(r, \theta)
\end{equation}
over the mean displacement data where $\rho$ is the global number density of cells. To directly produce a numeric value of $\left \langle d^2 \right\rangle^{mean}_{(m,a)}$ from the data, the integral is turned into a sum over all the bins in $r$ and $\theta$. The diffusion coefficient produced by these motions is then defined from Eq.~\ref{eq:app_affine_msd} as 
\begin{equation}\label{eq:app_affine_diffusion}
\mathrm{D}^{mean}_{(m,a)} = \frac{N_{cells}}{4 \tau_\delta} \left \langle d^2 \right\rangle^{mean}_{(m,a)}
\end{equation}

\begin{figure}[!t]  
\begin{center}
    \includegraphics[width=0.48\textwidth]{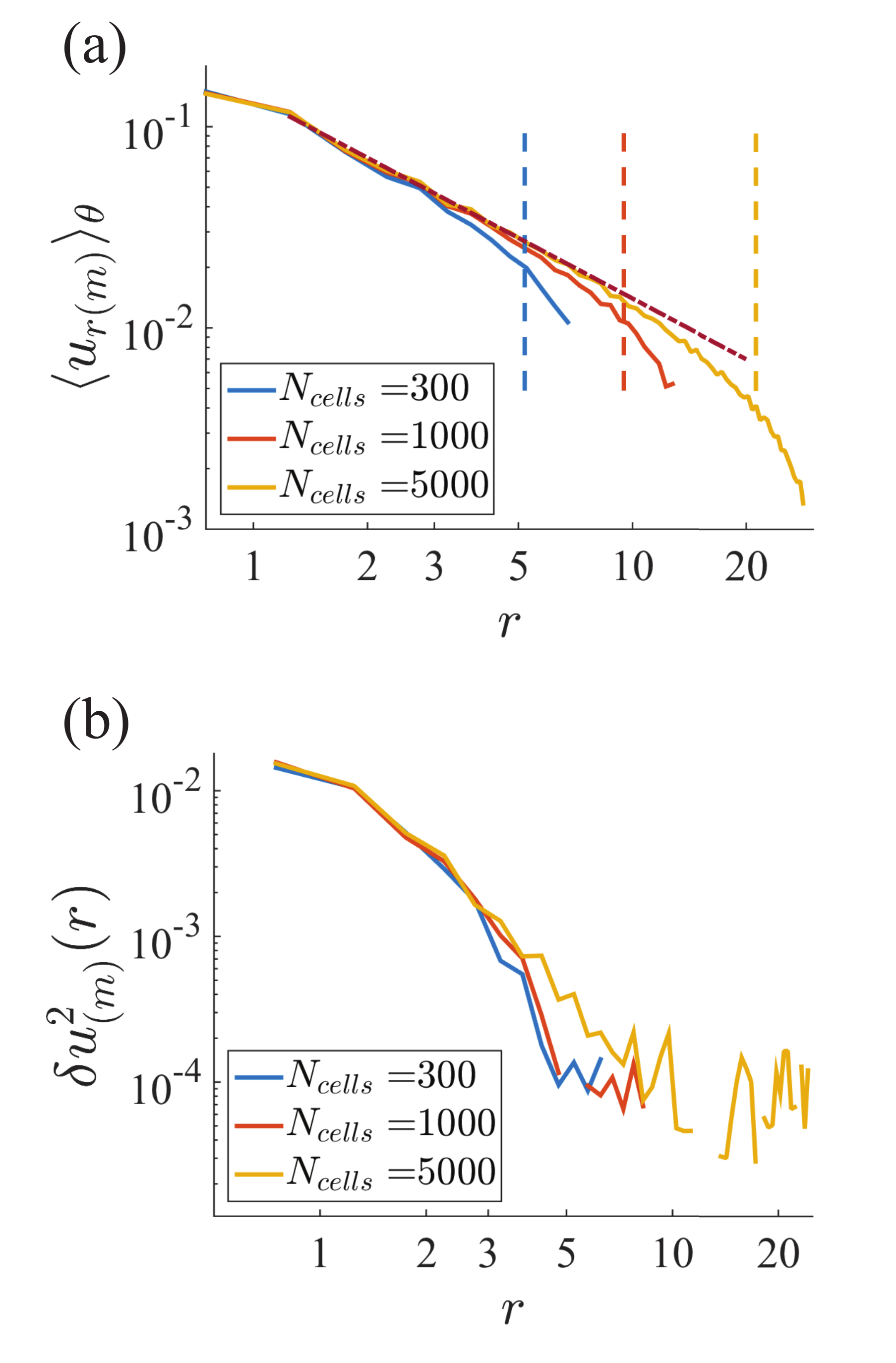}
    \caption{\label{fig:app_single_events} Scaling of mitosis-based displacement fields with system size for a tissue with $s_0=3.8, v_0=0.05, D_r = 1.0$. In \textbf{(a)} the \mdc{mean} displacements have consistent form of a $1/r$ field which fills up to a finite fraction of the box dimension. Dotted vertical lines indicate the cutoff estimate $r_{large} = 0.3 \sqrt{N_{cells}/\rho}$. In \textbf{(b)}, the \mdc{fluctuating} displacements extend over a finite region of fluidization and are unchanged by increasing system size.}
    \end{center}
\end{figure}

The estimate produced above leads to useful and accurate predictions of the dynamics as shown in Figures~\ref{fig:dd_diffusion_check} and \ref{fig:dd_collapses} in the main text. However, this prediction provides little insight into the functional form of these \mdc{mean} diffusions. Some insight can instead be gained by estimating the integral in Eq.~\ref{eq:app_displacement_averaging_affine} using some assumptions based on the data. In Fig.~\ref{fig:app_single_events}(a) we see that averaging these \mdc{mean} displacements over the angular bins reveals a $1/r$ trend which may be captured by  $ u_{(m,a)}(r) = u_{(m,a)}/(\rho r) $.  This trend persists out to a distance of $\sim 0.3 L$ where the event begins to interfere with its periodic image and drops off quickly (here $L = \sqrt{N_{cells}/\rho}$ is the periodic box length). A good approximation of the average displacement is given by
\begin{equation}\label{eq:dd_affine_integral}
\left \langle d^2 \right\rangle^{mean}_{(m,a)} = \frac{ 2 \pi \rho}{N_{cells}} \int_{r_{small}}^{r_{large}} \mathrm{d}r r\langle\mathbf{u}_{(m,a)}^2\rangle_{\theta}(r) \, ,
\end{equation}
where the small $r_{small}$ and large %$r_{large}=L_c\sqrt{\frac{N_{cells}}{\rho}}$ 
$r_{large} \sim  0.3 L$ cutoffs of integration respectively capture the typical cell neighbor spacing and the extent of the \mdc{mean} field. Inserting the above form of $ u_{(m,a)}(r) $ and integrating, we find an expression for the diffusion
\begin{equation}
\begin{aligned}\label{eq:app_toy_diff_affine}
\mathrm{D}^{mean}_{(m,a)} = \frac{N_{cells}}{4 \tau_\delta} & \langle\mathbf{u}^2_{(m,a)i} \rangle \\ & = \frac{ \pi u_{(m,a)}^2}{2 \tau_\delta \rho} ln\left( \frac{r_{large}}{r_{small}} \right) \, .
\end{aligned}
\end{equation}
\mdc{Inserting $r_{large} = 0.3 \sqrt{N_{cells}/\rho}$ we will find that this contribution to the diffusion scales as $\mathrm{D} \sim \log(N)$. However, this predicted scaling will only continue as long as the appropriate long-distance cutoff $r_{large}$ is set by the box size. As we consider increasingly large tissues, these \mdc{mean} fields will interfere with that from other events rather than periodic images of the same event. Therefore we expect that the cutoff length (and by proxy, the predicted $\mathrm{D}^{mean}_{(m,a)}$) will become independent of the tissue size in large enough tissues.}

%%%%%%%%%%%%%%%%%%%%%%%%%%%%%%%%%%%%%%%%%%%%%%%%%%%%%%%%%%%%%%%%%%%%%%%%%%%%%%%%%%%%%%
%%%%%%%%%%%%%%%%%%%%%%%%%%%%%%%%%%%%%%%%%%%%%%%%%%%%%%%%%%%%%%%%%%%%%%%%%%%%%%%%%%%%%%
%%%%%%%%%%%%%%%%%%%%%%%%%%%%%%%%%%%%%%%%%%%%%%%%%%%%%%%%%%%%%%%%%%%%%%%%%%%%%%%%%%%%%%

\subsection{Estimating \mdc{fluctuating} contributions to displacement} \label{sec:app_noise_estimate}
Similar to the previous section, we now consider the contribution to diffusion from the \mdc{fluctuating} (third and fourth) terms in Eq.~\ref{eq:dd_total_toy_diffusion}. 
Similar to Eq.~\ref{eq:app_affine_dynamics}, the contribution to the tracer dynamics generated by these \mdc{fluctuating} displacements is written as
\begin{equation}\label{eq:noise_dynamics}
\Delta\mathbf{r}_{(m,a)}^{fluc}(t) = \sum_{i = 0}^{n(t)} \delta \mathbf{u}_{(m,a)i} \, ,
\end{equation}
where $\delta \mathbf{u}_i$ is the \mdc{fluctuating} displacement produced by the $i$-th mitosis(apoptosis) event. This displacement is assumed to have a random direction and magnitude determined by the distribution of $\delta u_{(m,a)}(r)$ from Eq.~\ref{eq:dd_nonaffine_displacements}. The mean square displacement again relies on the quantity $\left \langle d^2 \right \rangle^{fluc}_{(m,a)}$, which again takes the form of a spatial average
\begin{equation}\label{eq:dd_nonaffine_integral}
\left \langle d^2 \right\rangle^{fluc}_{(m,a)} =\frac{\rho}{N_{cells}} \int \mathrm{d}^2 \mathbf{x} \, \delta \mathbf{u}^2_{(m,a)i}(r, \theta) \, .
\end{equation}
Turning this integral into a sum over bins, the \mdc{fluctuating} displacement data from Section~\ref{sec:dd_single_events} can be used to estimate these contributions to diffusion
\begin{equation}\label{eq:app_nonaffine_diffusion}
\mathrm{D}^{fluc}_{(m,a)} = \frac{N_{cells}}{4 \tau_\delta} \left \langle d^2 \right\rangle^{fluc}_{(m,a)} \, .
\end{equation}

Similar to the case of the \mdc{mean} displacements, we may estimate the form of the diffusion Eq.~\ref{eq:dd_nonaffine_integral} based on some more tangible parameters. This is accomplished by approximating the \mdc{fluctuating} displacements as a uniform region of fluidization with $\delta u_{(m,a)}(r) = w_{(m,a)}\Theta(R_w - r)$. The diffusion is then
\begin{equation}\label{eq:toy_diff_nonaffine}
  D^{fluc}_{(a,m)}   = \frac{\rho \pi w_{(m,a)}^2 R_{w}^2}{ 4 \tau_\delta} \, .
\end{equation}
Fig.~\ref{fig:dd_single_events} (b) indicates that all of these parameters above are independent of system size as long as $L$ is sufficiently larger than  $R_{w}$. Therefore the contribution of these displacements to cell dynamics is also independent of system size as long as $\sqrt{N_{cells}/\rho} > 2 R_{w}$.

These \mdc{fluctuating} components may also be checked for scaling with increasing system size. As shown in Fig.~\ref{fig:app_single_events}-b, the mitosis event creates a finite region of fluidization which is consistent across system sizes.

\bibliography{bibliofile}
\end{document}